\documentclass[twocolumn]{aastex63}
\usepackage{CJKutf8}
\usepackage{hyperref}
\usepackage{amsmath}
\usepackage{pbox}
\newcommand{\cntext}[1]{\begin{CJK}{UTF8}{gbsn}#1\end{CJK}\kern-1ex}

\newcommand{\uat}[2]{\href{http://astrothesaurus.org/uat/#2}{#1 (#2)}}
\shorttitle{Wei et al. 2024}
\shortauthors{Wei et al.}
\graphicspath{{./}{figures/}}
\turnoffeditone

\begin{document}

\title{Episodic energy release during the main- and post-impulsive phase of a solar flare}

\author[0000-0002-6628-6211]{Yuqian Wei}
\affiliation{Institute for Space Weather Sciences, New Jersey Institute of Technology, 323 Martin Luther King Blvd, Newark, NJ 07102-1982, USA}

\author[0000-0002-0660-3350]{Bin Chen (\cntext{陈彬})}
\affiliation{Institute for Space Weather Sciences, New Jersey Institute of Technology, 323 Martin Luther King Blvd, Newark, NJ 07102-1982, USA}

\author[0000-0003-2872-2614]{Sijie Yu (\cntext{余思捷})}
\affiliation{Institute for Space Weather Sciences, New Jersey Institute of Technology, 323 Martin Luther King Blvd, Newark, NJ 07102-1982, USA}

\author[0000-0002-5233-565X]{Haimin Wang}
\affiliation{Institute for Space Weather Sciences, New Jersey Institute of Technology, 323 Martin Luther King Blvd, Newark, NJ 07102-1982, USA}
\affiliation{Big Bear Solar Observatory, New Jersey Institute of Technology, 40386 North Shore Lane, Big Bear City, CA 92314-9672, USA}

\author{Yixian Zhang}
\affiliation{School of Physics \& Astronomy, University of Minnesota Twin Cities, Minneapolis, MN 55455, USA}

\author{Lindsay Glesener}
\affiliation{School of Physics \& Astronomy, University of Minnesota Twin Cities, Minneapolis, MN 55455, USA}

\correspondingauthor{Yuqian Wei}\email{yw633@njit.edu}

\begin{abstract}

When and where the magnetic field energy is released and converted in eruptive solar flares remains an outstanding topic in solar physics. To shed light on this question, here we report multi-wavelength observations of a C9.4-class eruptive limb flare that occurred on 2017 August 20. The flare, accompanied by a magnetic flux rope eruption and a white light coronal mass ejection, features three post-impulsive X-ray and microwave bursts immediately following its main impulsive phase.  For each burst, both microwave and X-ray imaging suggest that the non-thermal electrons are located in the above-the-loop-top region. Interestingly, contrary to many other flares, the peak flux of the three post-impulsive microwave and X-ray bursts shows an increase for later bursts. Spectral analysis reveals that the sources have a hardening spectral index, suggesting a more efficient electron acceleration into the later post-impulsive bursts. We observe a positive correlation between the acceleration of the magnetic flux rope and the non-thermal energy release during the post-impulsive bursts in the same event. Intriguingly, different from some other eruptive events, this correlation does not hold for the main impulse phase of this event, which we interpret as energy release due to the tether-cutting reconnection before the primary flux rope acceleration occurs. In addition, using footpoint brightenings at conjugate flare ribbons, a weakening reconnection guide field is inferred, which may also contribute to the hardening of the non-thermal electrons during the post-impulsive phase. 

\end{abstract}


\section{Introduction}\label{sec:intro} 
The relationship between the kinematics of coronal mass ejections (CMEs) and the corresponding flaring emissions is important in understanding how magnetic energy is released and subsequently converted into different forms of energy in solar eruptions \citep[e.g.][]{lin2000effects}. The close temporal correlation between flare X-ray emission and the early kinematics evolution of the erupting magnetic flux rope/filament during the main impulsive phase (MIP) has been widely observed \citep{zhangTemporalRelationshipCoronal2001, Maricic2007, temmer2008acceleration}.  As a direct indicator of the flare reconnection, the rate of the magnetic flux change inferred using the advancing flare ribbon at the photosphere, is found to show a temporal correlation with the acceleration of the associated filament eruption/CME \citep{Qiu2004, hu2014structures}. In a statistical study, close correlations are found between the acceleration of the erupting filaments and the magnetic flux change rate \citep{Jing2005}. More recently, by analyzing a large sample of CME-associated flare events, \citet{zhu2020how} confirmed the correlation between the peak filament/CME acceleration and the peak rate of magnetic flux change. In addition, they revealed a positive correlation between the total reconnected magnetic flux and the maximum CME velocity in events accompanied with fast CMEs ($>\!600\ \mathrm{km}\ \mathrm{s}^{-1}$). Such a correlation suggests that the magnetic energy release rate is closely related to flux rope eruption. The correlation is also revealed in resistive magnetohydrodynamic (MHD) simulations \citep{cheng2003fluxa, Reeves2006, reeves2010relating}.

As discussed by \citet{welsch2018flux}, the mechanisms of the flux rope/CME acceleration can be grouped into two general categories: (1) Lorentz force in and around the magnetic flux rope that directly drives its acceleration. These models suggest the increasing Lorentz force during the flare impulsive phase is attributed to the flare reconnection that adds poloidal magnetic flux to the flux rope and, at the same time, reduces the tension force from the overlaying constraining magnetic fields \citep{lin2000effects, lin2005direct}. (2) Acceleration of the flux rope that is due to momentum transferred from the upward-directed reconnection outflows. The highly bent post-reconnection field lines coming out of the diffusion region bear a large magnetic tension force and are accelerated to near the Alfvén speed \citep{parker1957sweet}. 
After joining the flux rope and becoming ``dipolarized,'' the upward-directed reconnection outflows transfer the momentum to the flux rope and facilitate its acceleration \citep{wang2007direct, xue2016observing, jiang2021}. 
Although multiple models have been proposed to account for the initiation and acceleration of the flux rope, such as the loss of equilibrium model \citep{lin2000effects}, tether-cutting reconnection model \citep{Moore2001}, breakout reconnection model \citep{antiochos1999model, karpen2012mechanisms}, 
the net change of the upward Lorentz force and/or the added momentum transfer from the upward reconnection outflows are both related to the magnetic flux change due to the ongoing flare reconnection. Hence, a positive correlation between the flux rope acceleration and flare energy release is expected.

The electron acceleration rate, implicated by the intensity of nonthermal hard X-ray (HXR) and/or microwave emission, is also found to be correlated to the rate of the flare energy release. By measuring the photospheric magnetic field and ribbon expansion, \citet{Qiu2004} inferred the rate of the magnetic flux change and reconnection electric field evolution during the impulsive phase of two two-ribbon flares, which are found to be temporally correlated with the microwave emission and the derivative of the SXR emission (as a proxy for HXR emission assuming the Neupert effect \citep{neupert1968comparison}). Adopting a similar method,  \citet{liu2009reconnection} analyzed 13 two-ribbon flares and found an anti-correlation between the average reconnection electric field and minimum overall photon spectral index from HXR observation. \citet{temmer2007energy} found that the local reconnection electric field inferred from the ribbon expansion in H$\alpha$/UV observations is uneven along the direction of the magnetic polarity inversion line (PIL). The locations of the spatially resolved HXR footpoint sources were found to be spatially correlated with the local electric field. With the observation of the flare ribbon from the Interface Region Imaging Spectrograph (IRIS; \citealt{depontieu2014interface}, \citet{naus2022correlated} also revealed the strong correlation between the local magnetic flux change rate and the production of the nonthermal electrons inferred from RHESSI HXR data. These studies suggest that the electron acceleration in flares and its temporal and spatial evolution are intimately related to the local electric field in the reconnection region.

Recent modeling studies have suggested that, in addition to the absolute magnetic energy release rate, the guide field, defined as the magnetic field component in the same direction as the reconnection current, also plays a key role in determining the efficiency of particle energization. In particular, a strong guide field can suppress the acceleration of electrons to high energies, resulting in an overall soft spectrum \citep{pritchett2004threedimensional, dahlin2014mechanisms,  dahlin2016parallel, li2017particle, arnold2021electron}. At present, direct means of measuring the guide field in the reconnection region has not been available, although the present microwave imaging spectroscopy observations, made possible by the Expanded Owens Valley Solar Array (EOVSA; \citealt{Gary2018}), have provided constraints of the guide field by comparing the overall magnetic field profile along the current sheet to model predictions \citep{chen2020measurement}\footnote{Prospects of achieving more direct measurements of the guide field can be realized by using microwave imaging spectropolarimetry.}. Alternatively, the inclination angle of the post reconnection flare arcade, usually constrained by comparing the orientation of the conjugate ribbon brightenings with respect to the magnetic PIL, can be used as a proxy to infer the guide field component in the coronal region \citep[e.g.][]{qiu2017elongationa, qiu2023role}. Recently, using three-dimensional MHD simulations, \citet{dahlin2022variability} revealed a prominent decrease of the guide field over the impulsive phase of eruptive flares, conforming to the observational evidence reported by \citet{qiu2010reconnection}.

However, the correlations discussed above have focused largely on the peak rates derived during the main impulsive phases of the flare-CME events and are based on statistical studies of multiple different events. In this work, we report a new finding that such a correlation is also present during the post-impulsive phases (PIP) of a single eruptive flare event. We also compare the difference in the geometry of the magnetic reconnection in the main- and post-impulsive flare phases and its possible impact on flux rope acceleration and electron energization. 

In Section~\ref{sec:obs_overview}, we provide an overview of the event observed in multiple wavelengths and present the early evolution of the erupting flux rope. In Section~\ref{sec:mw_hxr}, we present microwave and X-ray imaging spectroscopy observations and spectral analysis of the main- and post-impulsive phase bursts. In Section~\ref{sec:risingfluxrope}, we report measurements of the kinematics of the erupting magnetic flux rope. In Section~\ref{sec:disc}, we interpret the observational results and discuss the implications based on the correlation between the flux rope acceleration, electron acceleration, and flare emission during the main- and post-impulsive phase.

\section{Observations}\label{sec:obs}
\subsection{Event Overview}\label{sec:obs_overview}
\begin{figure*}[!ht]
\centering
\includegraphics[width=0.68\textwidth]{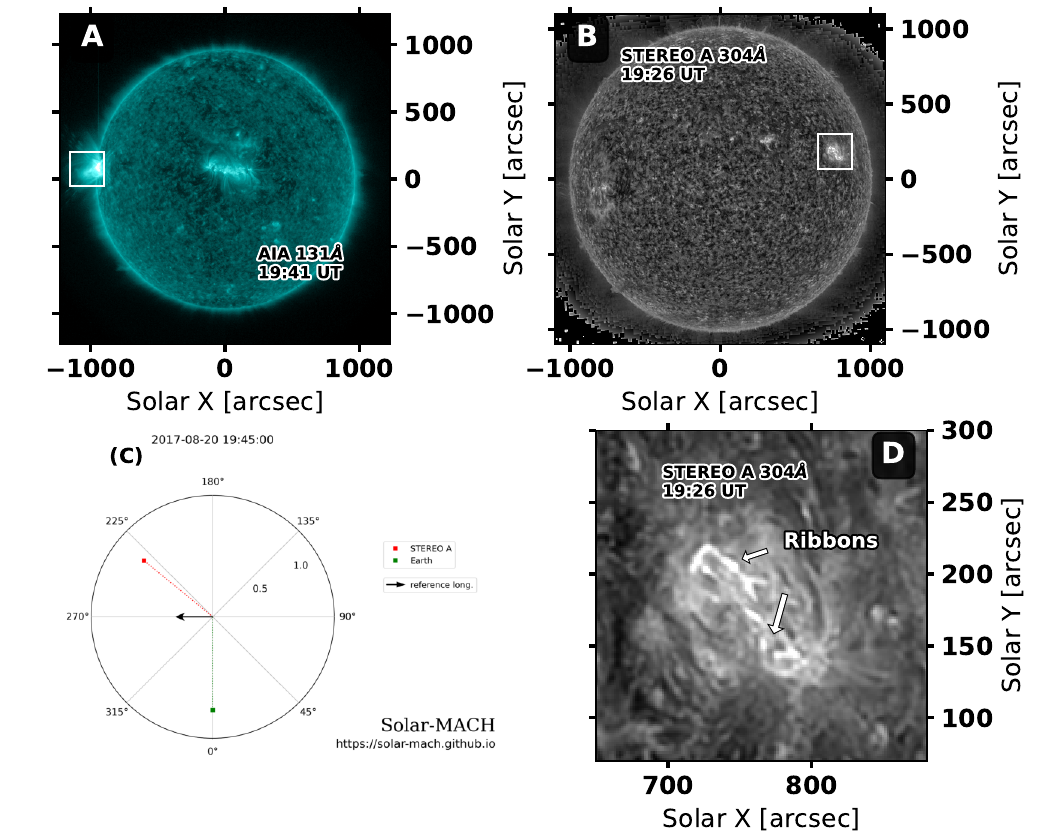}
\caption{\label{fig:overview_img} (a) The eruptive solar flare event under study as observed in EUV by the SDO/AIA 131~\AA\ filterband on 2017 August 20 at 19:41:00 UT. The white box shows the field of view (FOV) that is used in Figure~\ref{fig:small_fov}. (b) The event was observed in STEREO/EUVI 304~\AA\ at 19:26:00 UT, with an enlarged view shown in (d) (whose FOV is indicated by the white box in (b)). (c) Relative location of the STEREO-A spacecraft (red cube), the Earth/SDO spacecraft (green cube), and the longitudinal direction of the event in the frame of the HEE coordinate system (produced using the Solar-MACH software; \citealt{gieseler2023solarmach}).}
\end{figure*}

\begin{figure*}[!hb]
\centering
\includegraphics[width=0.8\textwidth]{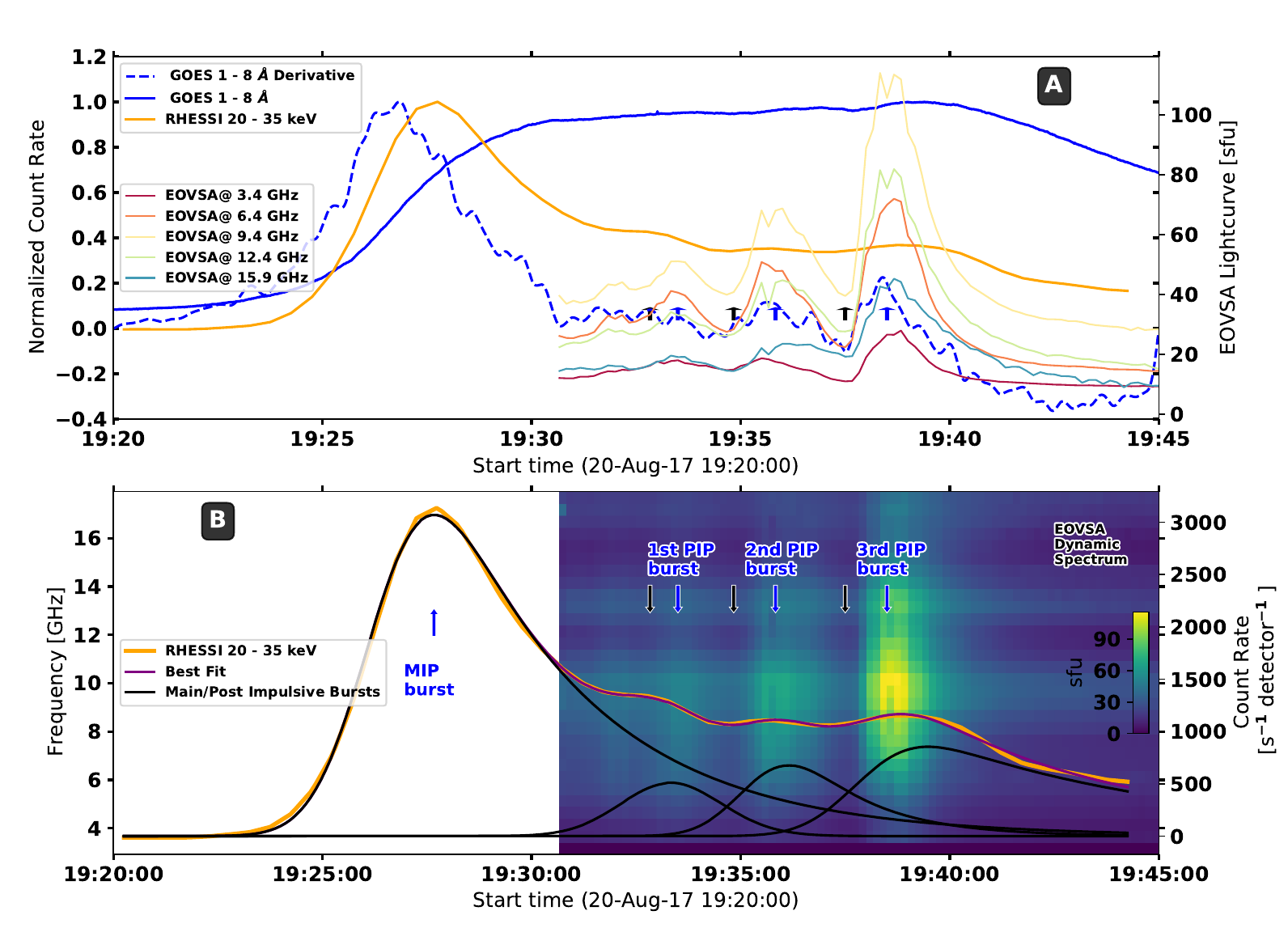}
\caption{\label{fig:overview_lc} (a) RHESSI 20--35 keV X-ray (orange curve), and GOES 1--8~\AA\ soft X-ray (SXR) (blue solid curve) light curves and its derivative (blue dashed curve) from 19:20 UT to 19:45 UT on 2017 August 20. Other color curves show EOVSA microwave flaring-region-integrated light curves at five selected frequencies (3.4, 6.4, 9.4, 12.4, 15.9 GHz) from 19:31 UT to 19:45 UT. The blue arrows indicate the peaks of the main-impulsive phase and the three post-impulsive phase bursts, while the black arrows indicate the corresponding pre-burst time of each burst used for background subtraction. (b) RHESSI 20--35 keV X-ray (solid orange curve) decomposed into the main-impulsive phase burst and the three post-impulsive phase bursts (solid black curves). The light curve is fitted using four components described in the text (solid purple curve). The background is the flare-region-integrated (same as that of the light curves in (a)) microwave dynamic spectrum from 19:31 UT to 19:45 UT.}
\end{figure*}

The C9.4-class event under study occurred on the east solar limb on 2017 August 20. The event was well observed in extreme ultraviolet (EUV) by the Atmospheric Imaging Assembly on board the Solar Dynamics Observatory (SDO/AIA; \citealt{pesnell2012, lemen2012}) and the Extreme UltraViolet Imager (EUVI; \citealt{wuelser2004euvi}) onboard STEREO-A, one of the two Solar Terrestrial Relations Observatory (STEREO; \citealt{kaiser2008stereo}) spacecraft. SDO/AIA observed the flare event near the east limb from the Earth's viewing perspective, while STEREO-A was $\sim$230$^{\circ}$ west from the Earth and provided observations from another vantage point (Figure~\ref{fig:overview_img}). The Geostationary Operational Environmental Satellite (GOES) and the Reuven Ramaty High-Energy Solar Spectroscopic Imager (RHESSI; \citealt{lin2002}) had full coverage of the event in X-rays. Meanwhile, EOVSA observed the post-impulsive phase and the decay phase of the event in the microwaves from 1--18 GHz (Figure~\ref{fig:overview_lc}). It missed the main impulsive peak because the antennas went off the Sun for calibration during that time.  

\begin{figure*}[!ht]
\centering
\includegraphics[width=0.9\textwidth]{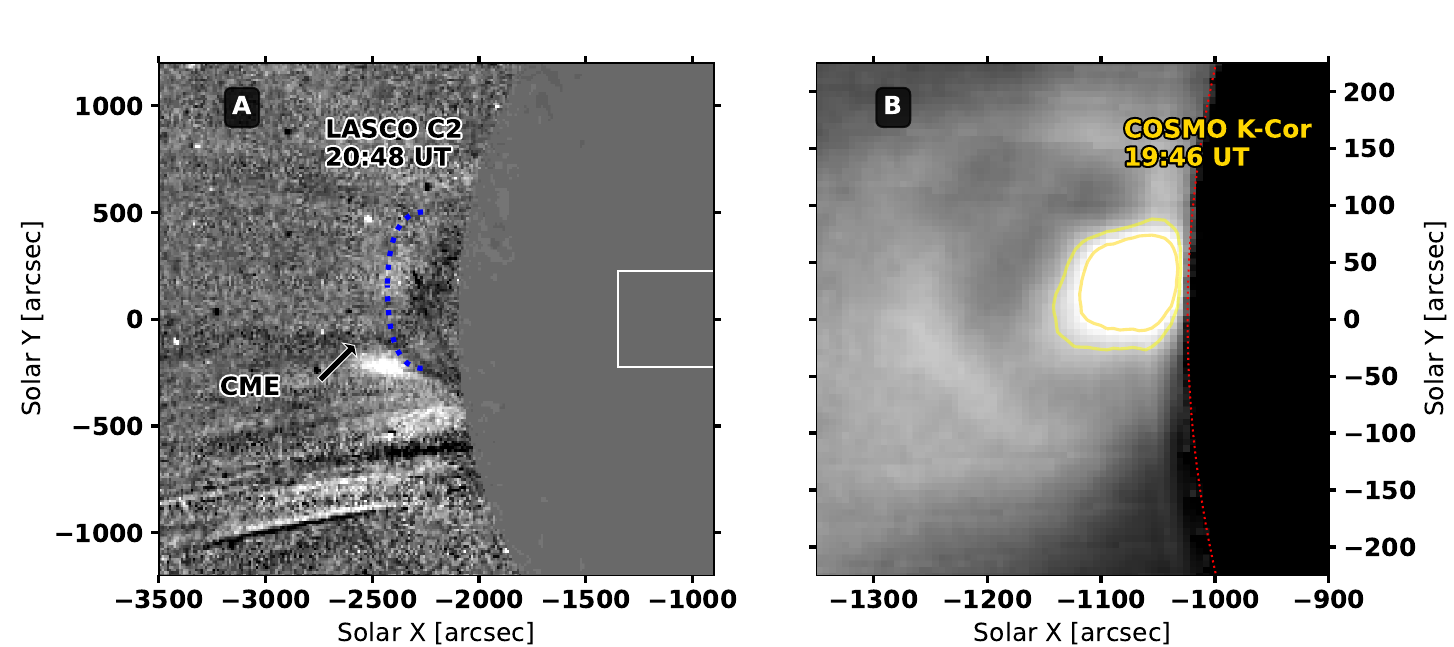}
\caption{\label{fig:lasco_kcor} The associated CME observed in white light by MLSO/K-cor and SOHO/LASCO C2. (a) The narrow CME as observed in LASCO/C2 running-difference image at 20:48 UT (81 minutes after the flare peak), whose front is indicated by the black arrow and the blue dashed curve. The white box shows the FOV of the MLSO/K-cor image in (b). (b) The three-part-structure CME as observed in MLSO/K-cor white light image at 19:46 UT (19 minutes after the flare peak). 
}
\end{figure*}

The event was associated with a white-light CME observed by the K-coronagraph of the Mauna Loa Solar Observatory (MLSO/K-Cor; \citealt{elmore2003calibration}), as well as the Large Angle Spectroscopic COronagraph on board the Solar and Heliospheric Observatory (SOHO/LASCO; \citealt{brueckner1995large}). Figure~\ref{fig:lasco_kcor}(a) shows a faint, slow ($\sim$$250\ \mathrm{km}\ \mathrm{s}^{-1}$), and narrow ($\sim$32$^{\circ}$) CME in the LASCO C2 difference image. In MLSO/K-Cor white light images, the CME, as shown in Figure~\ref{fig:lasco_kcor}(b), displays a typical three-part-structure: a leading front, a relatively darker cavity, and a bright core, which are typically explained as the plasma pileup at the leading edge of the erupting magnetic flux rope, the flux rope body, and the embedded filament/prominence, respectively \citep{illing1985observation, chen2011coronal, vourlidas2013how}, although certain studies suggested that the bright core may instead be attributed to the flux rope itself \citep{howard2017challenginga, veronig2018genesis,song2019nature}. In addition, another recent study, based on laboratory experiments, proposed an alternative model to interpret the cavity as a result of induced reverse current evacuating the background plasma away from the core \citep{2018ApJ...862L..15H}.

\begin{figure*}[!ht]
\centering
\includegraphics[width=0.9\textwidth]{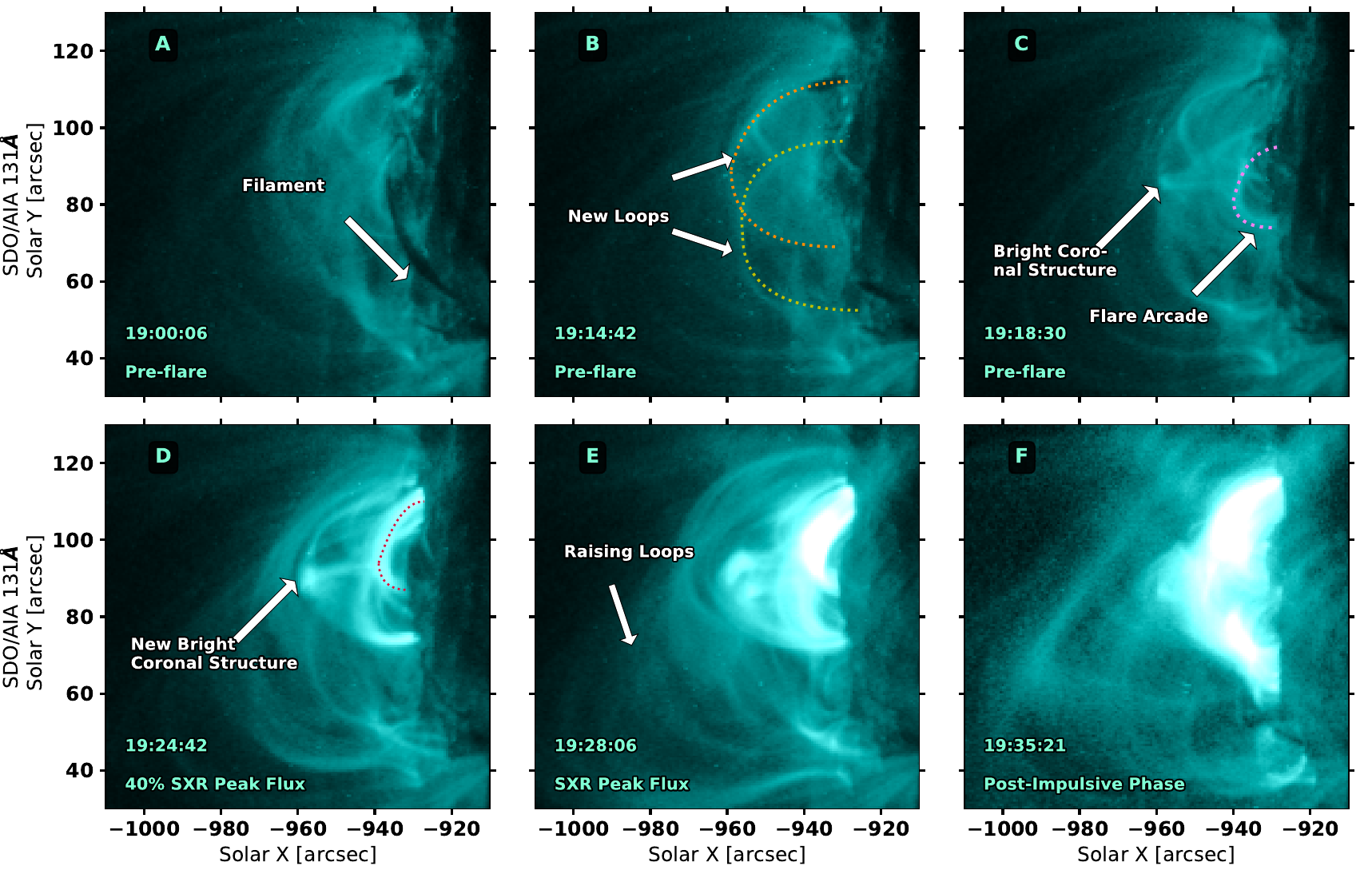}
\caption{\label{fig:small_fov} Close-up view of the coronal evolution as observed by SDO/AIA 131~\AA\ from the pre-flare phase to the beginning of the post-impulsive phase. (a) The dark filament which exists before the event, is marked by the white arrow. (b) The appearance of two loops during the pre-flare phase is highlighted by the yellow/orange dotted curves. (c) The newly formed coronal structure and the flare arcade (pink dashed curve) appear as the result of a tether-cutting reconnection during the pre-flare phase (indicated by the white arrows). (d) The gradually brightening coronal structure and a flare arcade appear during the main-impulsive phase, indicated by the white arrow and red dashed curve, respectively. (e) and (f) An Enlarged view of the flare region at the SXR flare peak and the beginning of the post-impulsive phase, respectively.}

\end{figure*}

Before the flare, a dark filament can be clearly identified in the EUV passbands (Figure~\ref{fig:small_fov}(a)). At 19:14 UT, two minutes before the HXR flux shows an early rise, two new loops appear in the SDO/AIA 131~\AA\ images (highlighted by the yellow/orange dotted lines in Figure~\ref{fig:small_fov}(b)). Four minutes later, at 19:18 UT, a flare arcade started to appear, which connected the two inner footpoints of the two loops (highlighted by the pink dotted lines in Figure~\ref{fig:small_fov}(c)).  Meanwhile, a new coronal structure also appears between the two loops (indicated by the white arrow in Figure~\ref{fig:small_fov}(c) and (d)). The event enters the main impulsive phase at 19:24 UT (c.f., Figure~\ref{fig:overview_lc}(a)), when the flare arcade further develops and brightens. The viewing perspective of the STEREO-A/EUVI reveals two J-shaped flare ribbons in a typical two-ribbon configuration (Figure~\ref{fig:overview_img}(d)).  
The bright coronal structure seen during the pre-impulsive phase disappeared when the event entered the post-impulsive phase (Figure~\ref{fig:small_fov}(f)).

\begin{figure*}[!ht]
\centering
\includegraphics[width=0.7\textwidth]{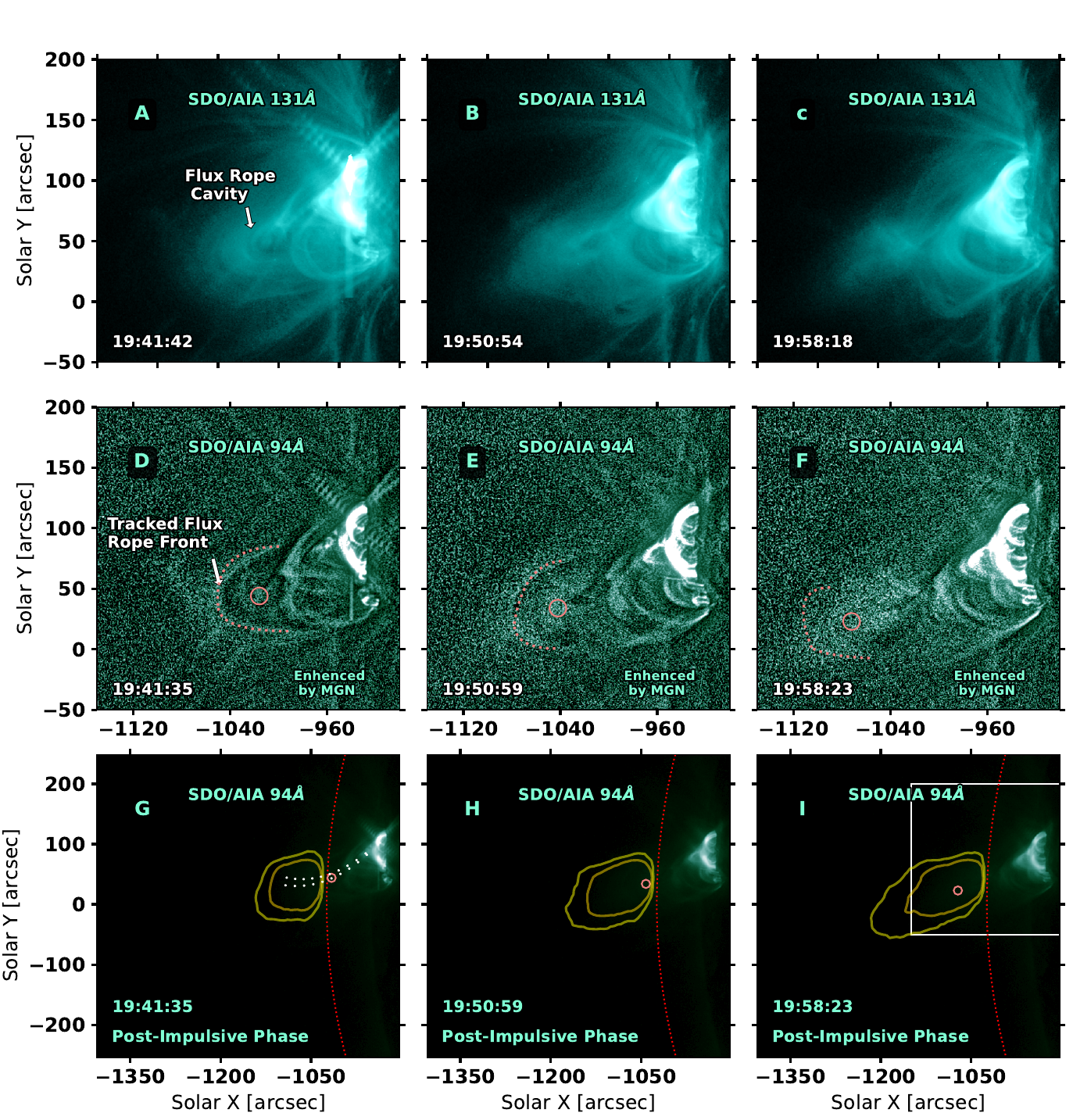}
\caption{\label{fig:big_fov} Evolution of the rising flux rope during the post-impulsive phase of the event as observed by SDO/AIA in EUV and MLSO/K-Cor in white light. (a)--(c) SDO/AIA 131~\AA\ images showing the evolution of the erupting flux rope cavity, indicated by the white arrow in (a). (d)--(f) SDO/AIA 94~\AA\ images enhanced with the MGN method. The overlying loop-like feature at the front of the cavity is indicated by the white arrow in (d) and the pink dashed curve in (d)--(f), while the center of the flux rope cavity is indicated by the pink open circle. See Animation~\ref{video:stack_plot} for the visualization of the rising overlying loop-like feature. (g)--(h) Contours of the MLSO/K-Cor white light images (the contour levels are same as that of Figure~\ref{fig:lasco_kcor}(b) ) overlaid on SDO/AIA 94~\AA\ images. The same pink open circles in (d)--(f) are also shown. The lower boundary of MLSO/K-Cor's FOV at 1.07 solar radii is indicated by the red dashed curve. The FOV that is used in (a)--(f) is shown as the white box in (i).}

\end{figure*}

As the event entered the main-impulsive phase, a group of large-scale overlying loops, which bridged the northern and southern ends of the active region, started to rise towards the southwest direction in succession (indicated by the white arrow in Figure~\ref{fig:small_fov}(e)). The eruption started to appear during the post-impulsive phase, when an oval-shaped cavity became visible in the SDO/AIA 131 \AA\ images ((Figure~\ref{fig:big_fov}(a))), which we interpret as the cross-section of the erupting magnetic flux rope.

To show the evolution of the erupting cavity more clearly, in Figures~\ref{fig:big_fov}(d)--(f), we show SDO/AIA 94 \AA\ images enhanced with the multi-scale Gaussian normalization (MGN) method \citep{Morgan2014}. A kernel size of 4$''$.8 is selected to sharpen the edge and reveal the loop-like structure at the front of the cavity. 
This loop-like structure will be used to measure the kinematics of the eruption, which will be discussed in detail in Section~\ref{sec:risingfluxrope}.

When the cavity, whose center is indicated by the pink circle in Figure~\ref{fig:big_fov}(D)--(I)), rises into the inner field of view (FOV) of the MLSO/K-Cor at 1.07 $R_{\odot}$ (red dashed line in Figure~\ref{fig:big_fov}(G)--(I)), it moves synchronously with the CME bright core seen in the MLSO/K-Cor white light images (the innermost contour in Figure~\ref{fig:big_fov}(h), (i)). The synchronized motion of the cavity center in SDO/AIA 94~\AA\ images and the bright core of the K-Cor CME corroborate our interpretation of the structure as an erupting magnetic flux rope.

\subsection{Microwave and X-ray Bursts During the Main- and Post-impulsive Phase}\label{sec:mw_hxr}

Shortly after the impulsive X-ray peak at 19:27 UT, the event entered its post-impulsive phase. Almost at the same time, EOVSA went back to the Sun at 19:31 UT and fully covered the post-impulsive phase.  In this phase, three broadband bursts can be observed in the EOVSA 1--18 GHz dynamic spectrum (Figure~\ref{fig:overview_lc}(b)), which peak at 19:33 UT, 19:35 UT, and 19:38 UT respectively. The dynamic spectrum is produced by integrating the total flux of the flaring region using images integrated from 19:32 UT to 19:45 UT with 134 frequencies in 2.5--18 GHz over 31 evenly spaced spectral windows (referred to as SPW 0 to SPW 30) and is conducted with a temporal resolution of 10-s. While post-impulsive phase microwave bursts have been reported in the literature \citep[e.g.][]{Yu2020, kou2022microwave}, this event shows an increase in the peak intensity of bursts that occur later. The peak flux density at, 9.4 GHz for example, increases from 50 sfu for the first burst to 114 sfu for the last one. 

As shown in Figure~\ref{fig:overview_lc}, the bursts also have a response in GOES 1--8~\AA\ SXR light curve derivative and RHESSI 20--35 keV. To investigate the relationship between the microwave and HXR bursts, we carried out a forward fitting on the RHESSI 20--35 keV light curve using a `heating-decay'' function following the method described in \citet{gryciuk2017flare}. For each burst, the time profile is prescribed as:

\begin{equation}
f(t)=\int_0^t g(t') h(t-t') \mathrm{d}t',
\label{heating_decay}
\end{equation}
which is the result of a Gaussian-shaped pulse $g(t)$ convoluted with an exponential decay term $h(t)$. They are, respectively, 
\begin{equation}
g(t)=g_0 \exp \left(-\left(t-t_0\right)^2 / 2\sigma^2\right)
\end{equation}
and
\begin{equation}
h(t)=\exp (-D t),
\end{equation}

where $g_0$, $t_0$, $\sigma$, and D are the parameterized amplitude, peak time, the standard deviation of the Gaussian function, and exponential decay coefficient, respectively.

The observed RHESSI 20--35 keV light curve is fitted with four pulses, which correspond to the main-impulsive peak and the three post-impulsive bursts, respectively. Similar to post-impulsive microwave bursts, RHESSI 20--35 keV X-ray bursts also show an increase in the peak intensity for later post-impulsive bursts, as shown in Figure~\ref{fig:overview_lc}. 

\subsubsection{X-ray Imaging and Spectral Analysis}
We reconstruct the RHESSI 6--12 keV images using the \texttt{CLEAN} algorithm \citep{Hurford2002} with a 40-s integration time, based on measurements from detectors 3, and 8. The X-ray sources are plotted as green open contours in Figure~\ref{fig:mw}(a), (c)--(e). During the main-impulsive phase (Figure~\ref{fig:mw}(a)),  three sources can be distinguished. The main source is located at the top of the flare arcade. The lower (western) source coincides with the southern footpoint of the arcade, while the upper (eastern) source coincides with the bright coronal structure discussed in Section~\ref{sec:obs_overview}. During the post-impulsive phase, the upper X-ray source quickly fades away following the rise of its EUV counterpart, while the looptop and the footpoint sources remain. The detailed spatial evolution of the HXR source during the post-impulsive phase will be further presented and discussed in Section~\ref{sec:source_motion}.
 
\begin{figure*}[!ht]
\centering
\includegraphics[width=1.0\textwidth]{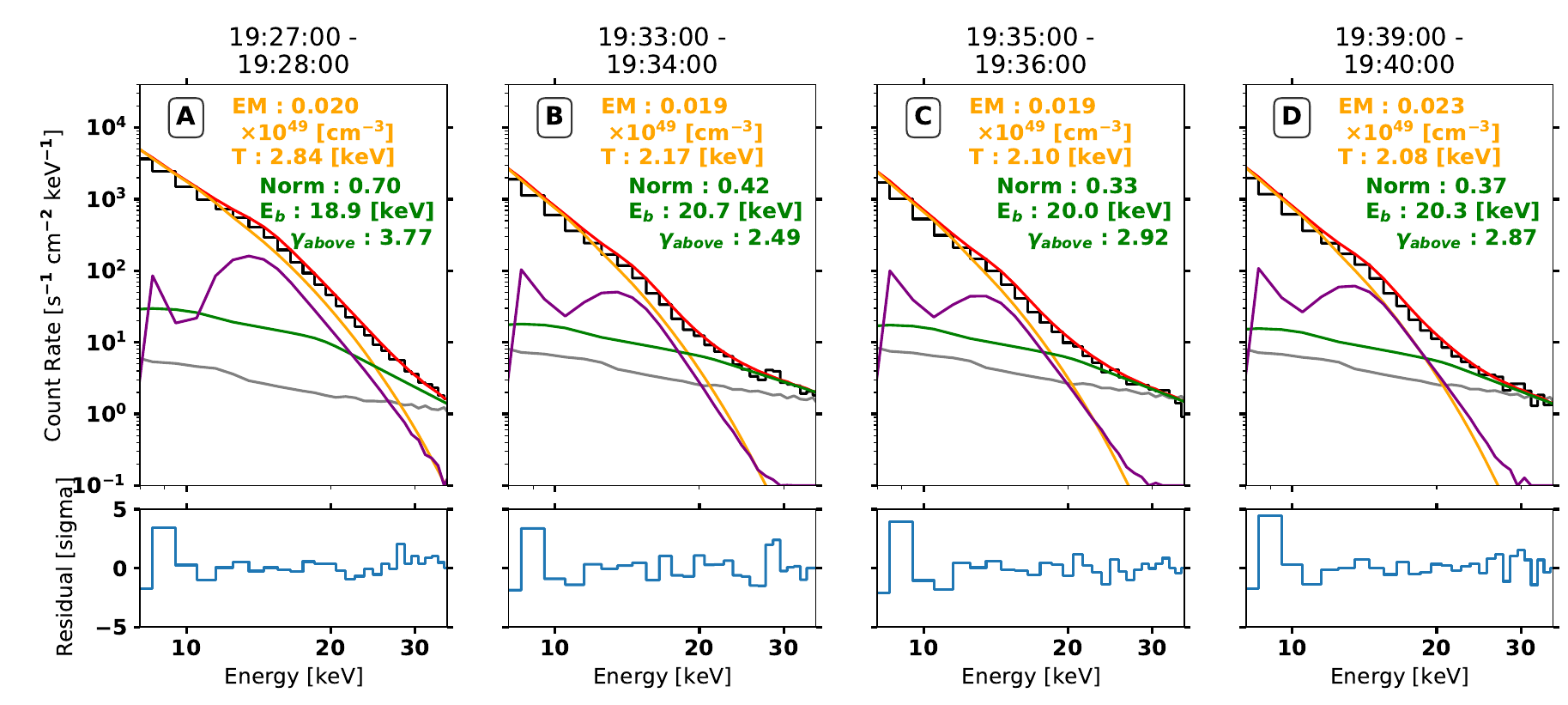}
\caption{\label{fig:rhessi_fitting} RHESSI photon flux spectra and spectral fitting results for the bursts during the main-impulsive phase (a) and the three post-impulsive phase burst (b)--(d). A single-temperature thermal bremsstrahlung model (\texttt{vth}, orange), a broken power-law model (\texttt{bpow}, green), and a pseudo-model to account for the pileup effect (\texttt{pileup\_mod}, purple) are fitted to the background-subtracted data (black). The background is depicted by the grey curve. The key fit parameters are summarized in Table~\ref{tab:rhessi_fitting_result}.}
\end{figure*}

We utilize the \texttt{OSPEX} tool, which is part of the SolarSoft IDL package (\texttt{sswidl}; \citealt{freeland1998data}) distribution, to perform the X-ray spectral analysis. Due to the increasingly more severe pulse pileup effect that affected RHESSI X-ray measurements toward the end of its operations, we limited our spectral analysis to data obtained from detector 3. This particular detector, thanks to its low sensitivity at the time of observation, showed the least amount of pileup effect among all active detectors. We note that the time profiles, especially those at the $>\!50\ \mathrm{keV}$ energy bands, have a time-varying, but smooth, background associated with each orbit. A polynomial fit to the time-varying background using the prior orbit is performed, which is subtracted from the time profiles of interest at all energies. After the background subtraction,  no detectable emission remains at $>\!50\ \mathrm{keV}$.
We fit three components to each X-ray pulse, which include a single-temperature thermal bremsstrahlung function \texttt{vth}, a broken power-law function \texttt{bpow}, and a pseudo function that accounts for the pileup effect \texttt{pileup\_mod} \footnote{A more detailed description of the pileup-correction module can be found at \url{https://sprg.ssl.berkeley.edu/~tohban/wiki/index.php/Pileup_mod_-_Pseudo_function_for_correcting_pileup}}, to the observed photon count rate spectrum. 
The fitting results for the X-ray pulses are presented and summarized in Table~\ref{tab:rhessi_fitting_result} and Figure~\ref{fig:rhessi_fitting}. The associated uncertainties are estimated using the built-in Monte Carlo module in \texttt{OSPEX}. 

\begin{table*}[!ht]
\centering

    \begin{tabular}{{ c | c |c | c | c | c |}}
   \toprule

& \pbox{2.5cm}{\textbf{Emission Measure [$10^{49}$$cm^{-3}$]}} 
& \pbox{2.5cm}{\textbf{Plasma Temperature [$keV$]}} 
& \pbox{2.5cm}{\textbf{Normalization at Epivot}} 
& \pbox{2.5cm}{\textbf{Break Energy [$keV$]}}
& \pbox{2.5cm}{\textbf{Negative Power-law Index}}\\
 
     \toprule
     \toprule
\textbf{main-impulsive Phase}&0.020$\pm$$5\mathrm{e}{-4}$ & 2.84$\pm$0.05  & 0.70$\pm$0.038& 18.9$\pm$0.5&3.77$\pm$0.25  \\
\textbf{1st PIP Impulse} & 0.019$\pm$$4\mathrm{e}{-4}$  & 2.17$\pm$0.05   &0.423$\pm$0.018& 20.7$\pm$2.4 &2.49$\pm$0.24\\
\textbf{2nd PIP Impulse} &  0.019$\pm$$2\mathrm{e}{-4}$ & 2.10$\pm$0.06   &0.328$\pm$0.010& 20.0$\pm$2.0&2.92$\pm$0.30\\
\textbf{3rd PIP Impulse} &  0.023$\pm$$3\mathrm{e}{-4}$ & 2.08$\pm$0.06  &0.366$\pm$0.008& 20.3$\pm$2.0&2.87$\pm$0.31\\
\end{tabular}
\caption{\label{tab:rhessi_fitting_result} RHESSI X-ray spectral fitting results at the main-impulsive phase and the three post-impulsive phase bursts. The 1-$\sigma$ uncertainties are estimated by running the built-in Monte Carlo analysis in OSPEX.}
\end{table*}

As shown in Figure~\ref{fig:rhessi_fitting}, the \texttt{bpow} component (yellow curves) represents a necessary contribution to the observed X-ray count spectra at energies above $\sim20\ \mathrm{keV}$ for each time period, signifying the presence of a nonthermal electron population throughout the main- and post-impulsive phase bursts. The nonthermal \texttt{bpow} component joins the thermal core at $\sim20\ \mathrm{keV}$ for all the four analyzed time intervals. Meanwhile, the background count rate dominates the spectra at $>\!35\ \mathrm{keV}$. Therefore, we integrate the 20--35 keV range to obtain the HXR light curve shown in Figure~\ref{fig:overview_lc}, which is used as a proxy for the time variation of the nonthermal component during the event. The power-law index of the observed X-ray photon spectrum above $\sim$20 keV (taken from \texttt{bpow}) decreases from $3.77$ during the main-impulsive phase to less than $3$ during the post-impulsive phase, suggesting a notably harder HXR photon spectrum in the post-impulsive phase. 

\subsubsection{Microwave Imaging Spectroscopy}
\begin{figure*}[!ht]
\centering
\includegraphics[width=0.8\textwidth]{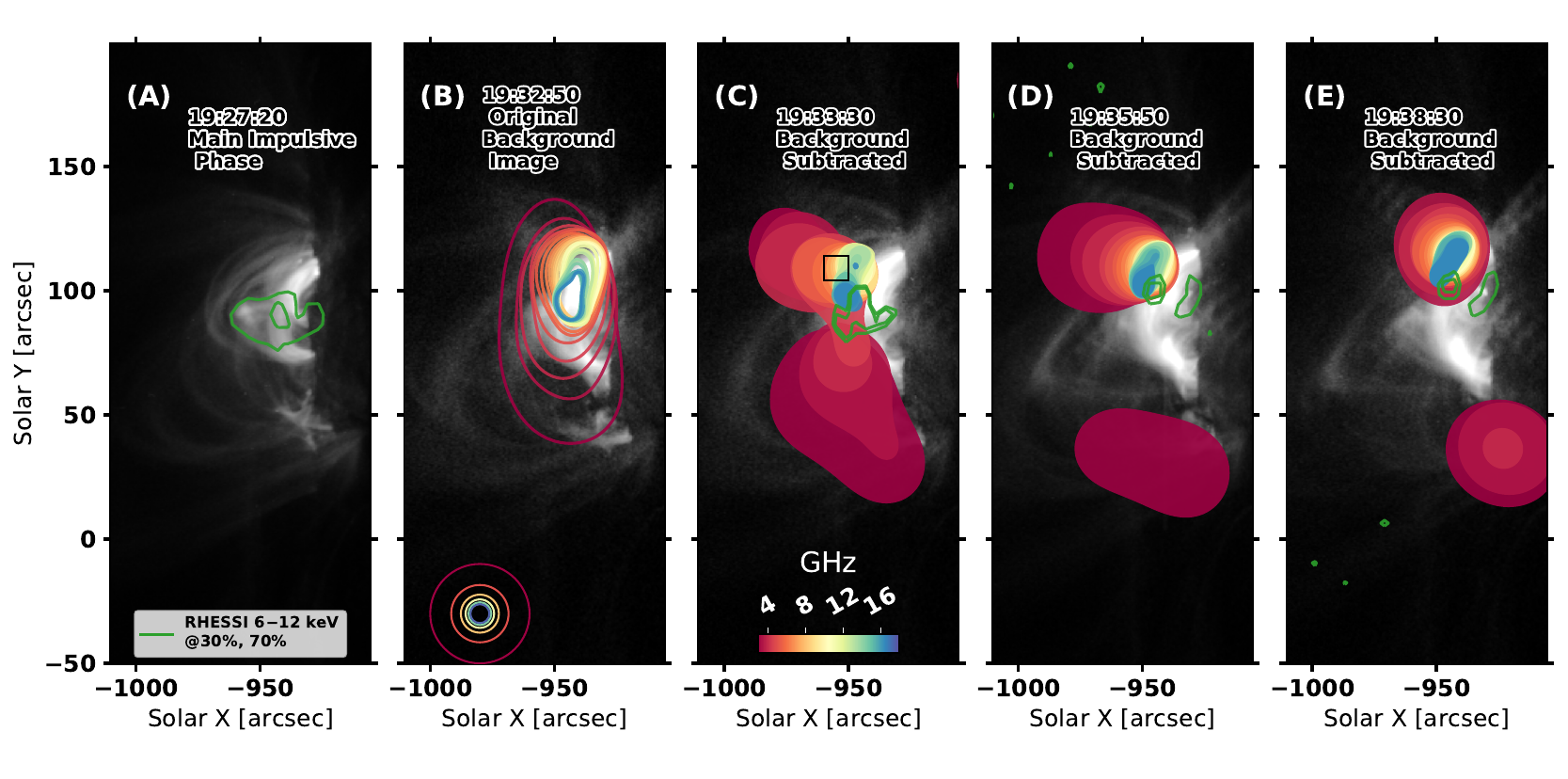}
\caption{\label{fig:mw} 
EOVSA multi-frequency microwave images and RHESSI 6--12 keV sources during the main impulsive phase and the three post-impulsive bursts. (a) RHESSI 6--12 keV source at the peak of the main-impulsive phase. The contour levels are 30\% and 70\% of the maximum. (b) Multi-frequency microwave images at the background time just prior to the first post-impulsive burst (the time is indicated by the first black arrow in Figure~\ref{fig:overview_lc}). The open contours are at the level of 50\% of the maximum brightness at each frequency. The restoring beam size of each frequency is shown in the bottom left corner. The microwave source morphology at all the selected background times prior to the three post-impulsive bursts is similar to each other. (c)--(e)  Background-subtracted multi-frequency microwave images at the peak of the three post-impulsive bursts (indicated by the three blue arrows in Figure~\ref{fig:overview_lc}). The level of the filled contours is as same as that of open contours in (b). The background gray-scale images are from SDO/AIA 131~\AA\ at the corresponding times. }
\end{figure*}

\begin{figure*}[!ht]
\centering
\includegraphics[width=0.8\textwidth]{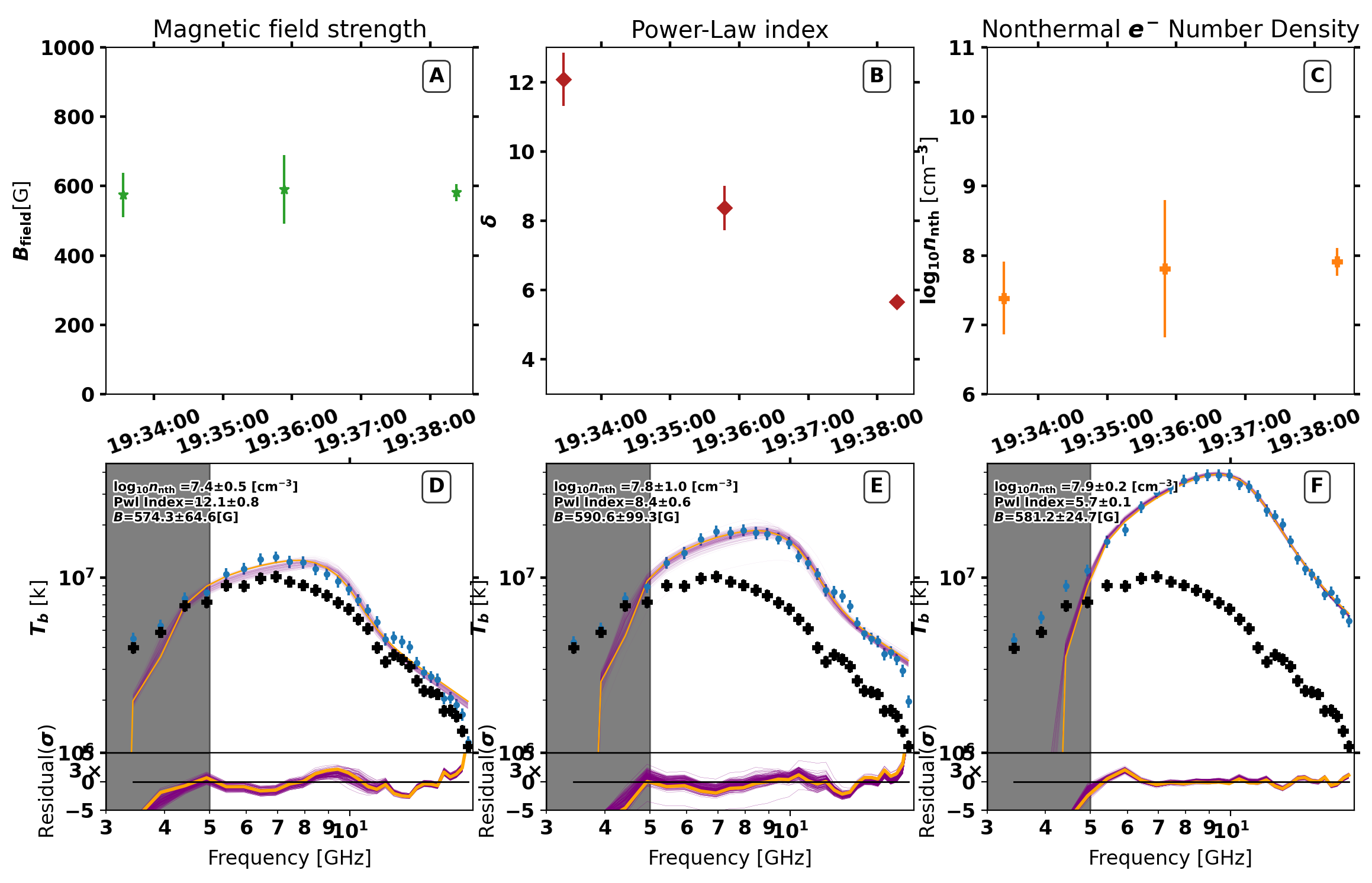}
\caption{\label{fig:spectral_fitting} Microwave spectra derived from the above-the-loop-top (ALT) region during the three post-impulsive bursts and corresponding spectral fitting results. (a)--(c) The evolution of the best-fit values of magnetic field strength $B$, power-law index of the electron energy distribution $\delta'$, and total nonthermal electron density $n_{\rm nth}$ above 20 keV, respectively. (d)--(f) EOVSA microwave brightness temperature spectra (blue dots) at the above-the-loop-top region during the post-impulsive phase bursts. The solid orange curve shows the best-fit results while the solid purple curve is the distribution of the MCMC runs within 1-$\sigma$ of the median MCMC values. The corresponding residuals are shown at the bottom of each panel.}
\end{figure*}

EOVSA missed the main-impulsive phase of the event but had full coverage of the post-impulse phase. Figure~\ref{fig:mw}(b) shows EOVSA microwave images as open contours (50\% of the maximum brightness at each frequency) at 19:32 UT just before the first post-impulsive phase burst. 
At high frequencies, the microwave source is concentrated on the northern part of the flare arcade. 
The sources extend from north to south (southeast in the viewing perspective of STEREO-A/EUVI images) and align with the ridge of the post-flare arcades. 
The microwave sources evolved rapidly during the post-impulsive phase. To better reveal their evolution, we performed difference imaging against the pre-burst background time in the visibility domain, with the selected background times pointed by the black arrows in Figure~\ref{fig:overview_lc}. The difference imaging results for the peaks of the three post-impulsive bursts (blue arrows in Figure~\ref{fig:overview_lc}) are shown in Figures~\ref{fig:mw}(c), (d), and (e), respectively. 
We verified that the sources we used as the background are stable over time, which show very similar morphology as that shown in Figure~\ref{fig:mw}(b) as well as nearly uniform spectral properties (black symbols in Figure~\ref{fig:spectral_fitting}. 

The background-subtracted microwave sources during three post-impulsive bursts are mainly located above the top of the bright post-flare arcades seen in SDO/AIA 131 \AA. For bursts 2 and 3, the above-the-looptop microwave source shows an obvious dispersion in height, with the high-frequency sources located at lower heights than the low-frequency ones. Such a microwave source morphology has already been reported by \citet{Gary2018,chen2020microwave} for the early impulsive phase of the X8.2 solar flare on September 10, 2017. The source dispersion above the looptop was interpreted as the signature of nonthermal electrons distributing along the reconnection current sheet, with the higher frequency source generally originating from sources regions with a greater magnetic field strength. In this event, however, we do not see a clear plasma sheet above the flare arcade, yet the physical origin of the frequency-dependent distribution of the source could be similar. During the peak of the three post-impulsive bursts,  a secondary source, only seen at low frequencies ($<$5~GHz), appears to the south of the above-the-looptop source. By comparing to the EUV images, we attribute the secondary microwave source as emission from the southern footpoint of the erupting flux rope, similar to the interpretation in \citet{chen2020microwave}. We will return to the discussion of the multi-wavelength flare context in Section~\ref{sec:disc}.
Figure~\ref{fig:spectral_fitting}(d)--(f) shows the brightness temperature spectra derived from the above-loop-top (ALT) region (black box in Figure~\ref{fig:mw}(c)) at the peak time of each post-impulsive phase burst (blue dots with error bars). The spectra show characteristics of the nonthermal gyrosynchrontron radiation \citep{dulk1985radio}. We fit the observed spectra using a nonthermal gyrosynchrontron radiation model from a homogeneous source with a power-law electron energy distribution based on the fast gyrosynchrotron codes developed by \citep{Fleishman2010}. Three free parameters are used in the spectral fitting: magnetic field strength $B$, the total number density of nonthermal electrons $n_{\rm nth}$, and the power-law index $\delta'$ of the nonthermal electron energy distribution ($f(\varepsilon)=dn_{\rm nth}(\varepsilon)/d\varepsilon\propto\varepsilon^{-\delta'})$. The energy range of the nonthermal electron distribution is fixed to 20 keV--10 MeV. The thermal electron density $n_{\rm th}$ and the plasma temperature $T$ is fixed to $1.7\times10^{11}~\mathrm{cm}^{-3}$ and 13 MK, respectively, which are based on the differential emission measure (DEM) analysis results within the same region using data from six SDO/AIA EUV channels (94, 131, 171, 193, 211, and 335~\AA). For the DEM analysis, we adopt the technique developed by \citet{hannah2012differential}. 
The column depth is assumed to be $13''$ based on the observed source size on the plane of the sky at 6.4~GHz. 

The best-fit spectra and the corresponding residual are shown as the thick solid orange curves in Figures~\ref{fig:spectral_fitting}(d)--(f). Following \citet{chen2020measurement}, we adopt the Markov Chain Monte Carlo (MCMC) method to estimate the uncertainties of the parameters and verify that we have achieved global minimization in the multi-dimensional parameter space. The spectra calculated from individual MCMC samples and the corresponding residuals are shown as thin purple curves in Figures~\ref{fig:spectral_fitting}(d)--(f). Figures~\ref{fig:spectral_fitting}(a)--(c) show the temporal evolution of the three fit parameters at the peak time of each post-impulsive phase microwave burst. The corresponding errors are estimated by the MCMC method. Of the three fit parameters, the only one that undergoes a significant change (against the uncertainties) is the power-law index of the nonthermal electron energy distribution $\delta'$. The power-law index shows a notable hardening for later post-impulsive microwave bursts, which increases from $\sim12$ for burst \#1 to $\sim5.5$ for burst \#3.

The brightness temperature spectra derived from the selected background times prior to the bursts are plotted as the black crosses in Figure~\ref{fig:spectral_fitting}(a)--(c). Spectral analysis indicates that the source is dominated by thermal emission. This thermal loop-top source is very similar to that reported by \citet{Yu2020} for the post-impulsive phase of the X8.2 solar flare on September 10, 2017. 

\subsection{Magnetic Flux Rope Kinematics}\label{sec:risingfluxrope}

To accurately track the kinematics of the erupting magnetic flux rope seen in SDO/AIA images in the low corona and the associated white-light CME observed in MLSO/K-Cor images, we produce a time-distance stack plot by obtaining the intensity along a slice shown as the dashed white lines in Figure~\ref{fig:big_fov}(g). The slice crosses the boundary of the FOV of SDO/AIA and extends to that of the MLSO/K-Cor. The eruption changes its course slightly during its ascent. Hence the slice is slightly bent toward the direction of the flux rope eruption. The width of the slice, defined as the perpendicular distance across the slit varying along the curved trajectory increases linearly from $3''$ at the bottom to $13.8''$ at the top. This adjustment compensates for the general expansion of coronal structures. In order to clearly show the dynamic features, we also apply the running-differential method on the SDO/AIA 94~\AA\ and MLSO/K-Cor white light images and then enhance the edge of those features with a high-pass filter technique. 

\begin{figure*}[!ht]
\centering
\includegraphics[width=0.9\textwidth]{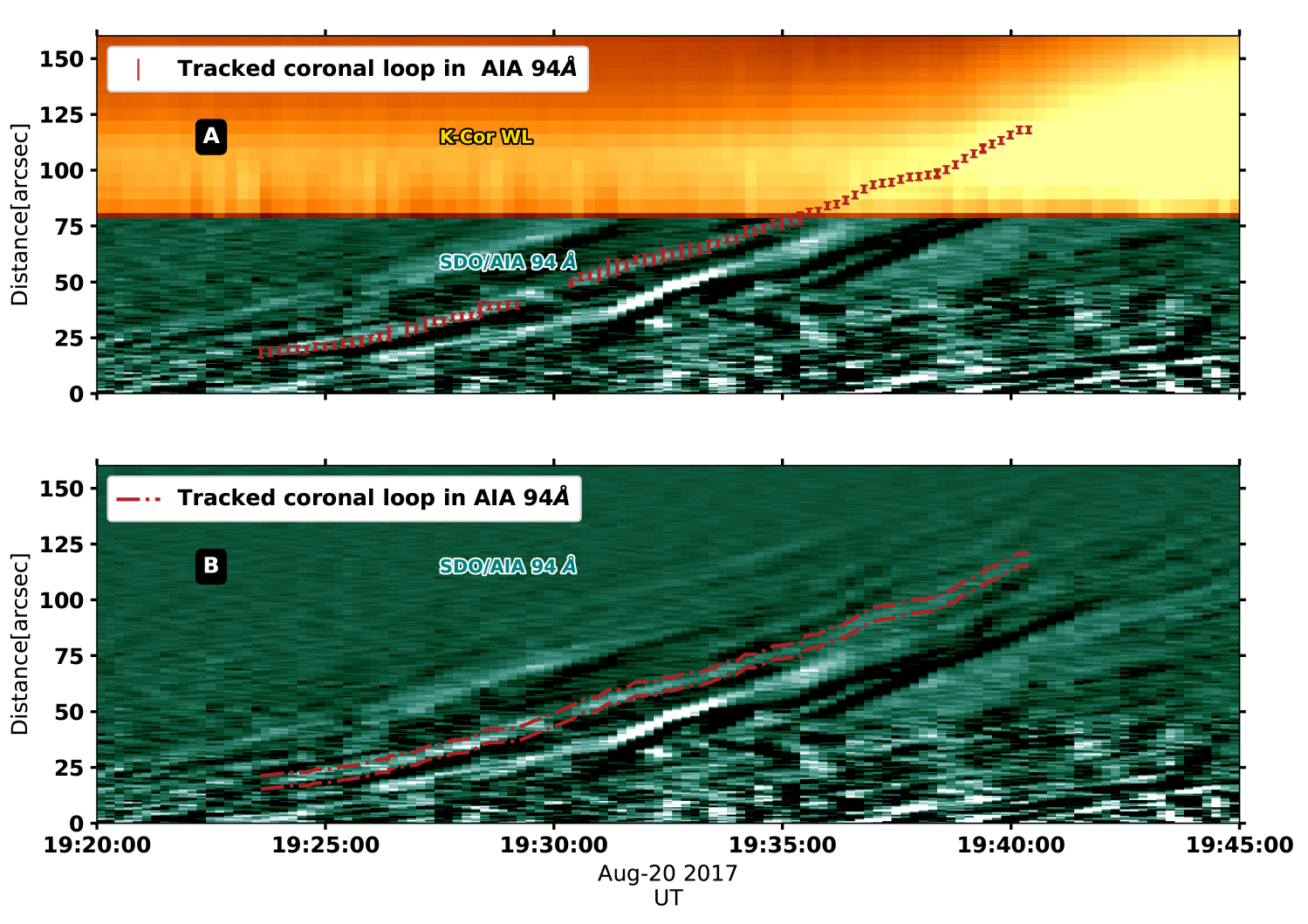}
\caption{\label{fig:stack_plot} (a) Composite time–distance stack plot of MLSO/K-Cor white light (upper) and SDO/AIA 94~\AA\ (lower; background-detrended) images series made along a slice as indicated by the dashed white curve in Figure~\ref{fig:big_fov}(g). Red symbols indicate the tracked eruption feature that represents the flux rope front in SDO/AIA 94~\AA\ images. The vertical lengths of the symbols indicate the corresponding uncertainties. (b) Same as (a) but shows the complete time–distance stack plot of the SDO/AIA 94~\AA\ image series. The trajectory of the tracked feature is sandwiched between the red dashed-dotted lines. An accompanying video (Animation~\ref{video:stack_plot}) shows the tracked feature on the SDO/AIA 94~\AA\ running-differential images and the corresponding time–distance plot.}
\end{figure*}

\begin{figure*}[!ht]
\centering
\includegraphics[width=0.9\textwidth]{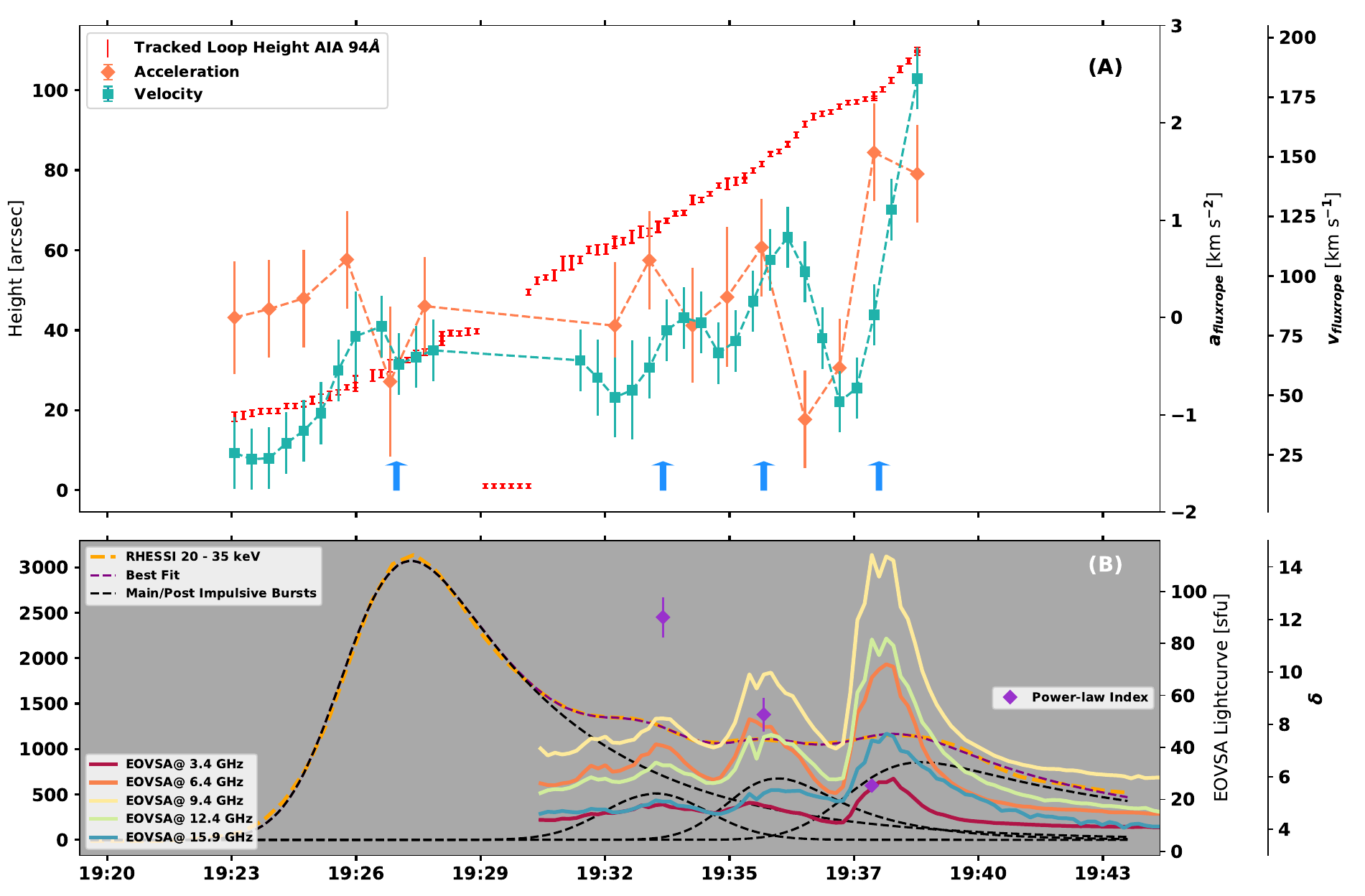}
\caption{\label{fig:traj} (a) Evolution of the height (red symbols), velocity (green symbols), and acceleration (orange symbols) of the erupting flux rope during the main- and post-impulsive phase. The peak time of the main- and post-impulsive phase bursts are indicated by the blue arrows. 
(b) Microwave and X-ray light curves during the same period (similar to Figure~\ref{fig:overview_lc}(b) but with the EOVSA microwave light curves shown instead). Also shown are the microwave-constrained power-law index of the nonthermal electron distribution at the peaks of the three post-impulsive phase bursts (purple symbols). 
 }
\end{figure*}

The resulting time-distance stack plot is shown in Figure~\ref{fig:stack_plot}. A number of rising tracks are visible at different heights, which correspond to not only structures that correspond to the core of the erupting magnetic flux rope, but also the overlying loops that enclose the flux rope cavity. 
We select one of the most visible tracks at the immediate front of the erupting cavity-like structure to measure the kinematics of the flux rope. The selected track is denoted by the red symbols in Figure~\ref{fig:stack_plot}(a) and the corresponding feature in the difference images is marked by pink dashed curves in Figure~\ref{fig:big_fov}(d)--(f). At any given time, the height of the tracked feature is determined by finding the peak in the intensity--height profile. To estimate the uncertainty of the tracked trajectory, we fit a skewed Gaussian function to the intensity--height profile around the peak. The full width at half maximum of this Gaussian function is used as an estimate of the uncertainty, shown as the vertical extension of the red symbols in Figure~\ref{fig:traj}(a). 

A similar time-distance stack plot is derived using the MLSO/K-Cor images at the same selected slice, shown in the upper portion of Figure~\ref{fig:traj}(a). It can be seen that the trajectories of the cavity front tracked in the SDO/AIA images and the upper edge of the white light CME core seen by MLSO/K-Cor are well aligned, which further demonstrates that we are tracking a coherent flux rope structure erupting from the low to middle corona. 

The speed and acceleration are derived from the height--time profile (red symbols), which are denoted by the light green and light orange symbols in Figure~\ref{fig:traj}(a), respectively. 

The speed and acceleration are derived from the height--time profile (red symbols), which are denoted by the light green and light orange symbols in Figure~\ref{fig:traj}(a), respectively. In our methodology, the derivatives are smoothed through rebinning after each derivation step, equivalently reducing the temporal resolution by a factor of two. When estimating the error in the first- and second-order derivative (velocity and acceleration), in addition to the error propagated from the height measurement itself, we include the smoothing error which is the root square (RMS) of the difference between the original profile and its smoothed counterpart, indicating the average smoothing error in both derivation steps.


Despite having a relatively large uncertainty, the acceleration of the flux rope in this event extends to and peaks at the post-impulsive phase.  
In particular, the last post-impulsive burst around 19:37 UT, which features the strongest microwave/X-ray flux and hardest electron energy spectrum (purple symbols in Figure~\ref{fig:traj}(b) and red symbols in Figure~\ref{fig:spectral_fitting}(b)), corresponds to the largest peak acceleration value. 
However, the relation in the main-impulsive phase does not appear to follow the same trend as in the post-impulsive phase. The maximum count rate of the RHESSI 6--20 keV time profile at the main-impulsive phase is 16 times larger than that at the first post-impulsive phase, while the corresponding maximum acceleration values are similar.

\subsection{Source motion at flare looptops and footpoints}\label{sec:source_motion}
\begin{figure*}[!ht]
\centering
\includegraphics[width=0.9\textwidth]{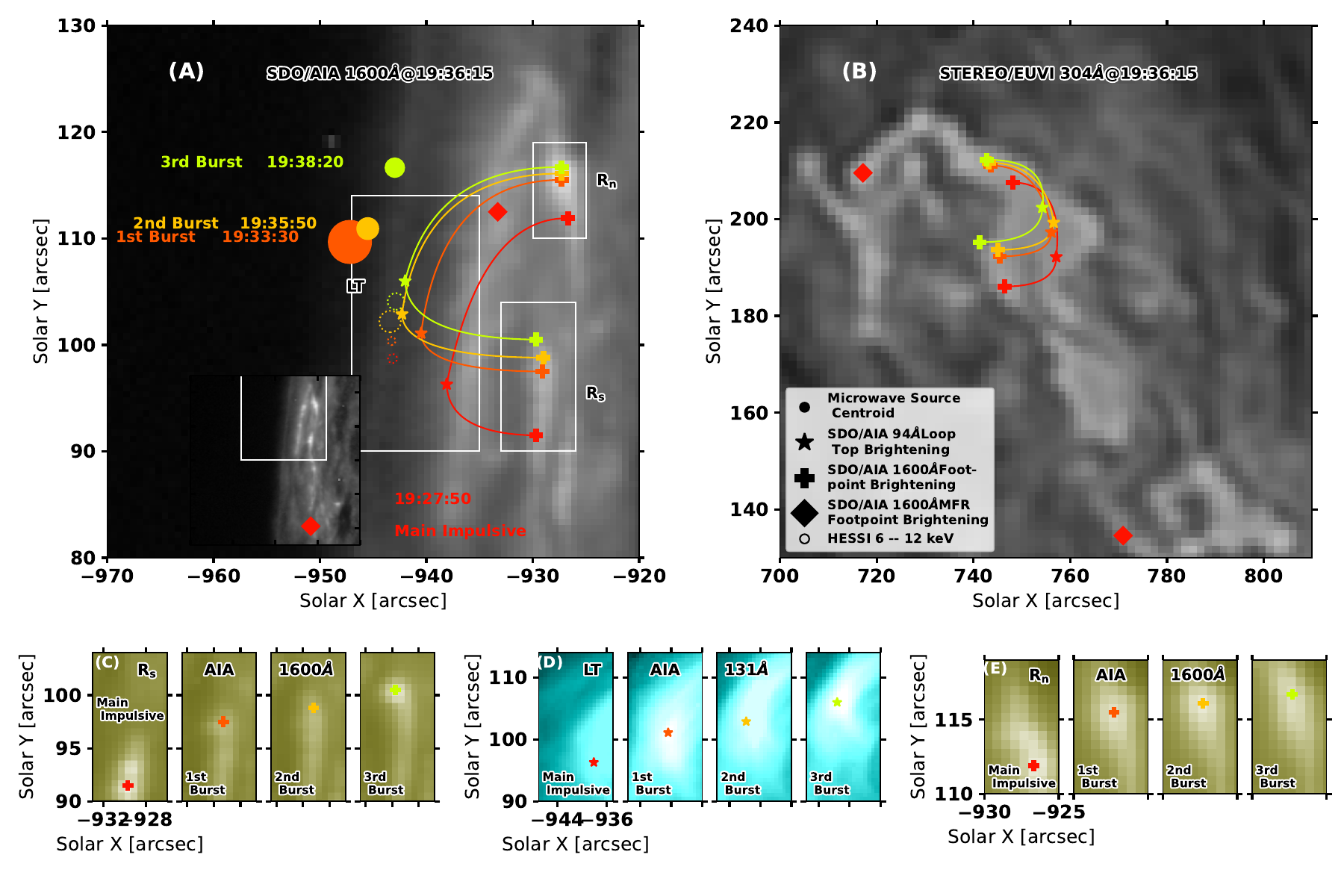}
\caption{\label{fig:source_motion} Synchronized northward motion of the microwave/X-ray/EUV looptop sources and corresponding UV footpoint brightenings. (a) Centroid locations of background-subtracted EOVSA microwave sources (filled circle), RHESSI 6--12 keV X-ray looptop sources (dashed open circle), EUV looptop brightening observed by SDO/AIA 131~\AA\ images(stars). Also shown are the northern ($R_{n}$)/ southern ($R_{s}$) ribbon brightenings in SDO/AIA 1600~\AA\  UV images (crosses) at the peak time of three post-impulsive bursts. The symbols are color-coded by time, as written in (a). The uncertainties of the microwave and X-ray sources are indicated by the radius of the corresponding filled/open circle. (b) Same as (a), but all the measured locations in (a) are re-projected into STEREO-A/EUVI's viewing perspective. (c)--(e) Detailed view of the northern ribbon region ($R_{n}$; c), flare looptop (LT; d), and southern ribbon region ($R_{s}$; e) at the peak time of the main impulsive phase and the three post-impulsive bursts. The corresponding FOV are indicated by the white boxes in (a).}
\end{figure*}

During the post-impulsive phase, we also observed a synchronized motion of energy release signatures above the flare looptops and at the footpoints. First, as shown in Figures~\ref{fig:mw}(c)--(e), despite the dispersion in height (largely in the east--west direction) as a function of frequency, the high-frequency background-subtracted microwave sources display an evident systematic motion from the southern to northern side of the flare arcade during the post-impulsive bursts. To better demonstrate the overall motion of the microwave source, in
Figure~\ref{fig:source_motion}(a), we show the centroid locations of the corresponding microwave sources as filled color circles. The centroid locations at each frequency are estimated by fitting the corresponding background-subtracted microwave source to a two-dimensional Gaussian ellipse using CASA task \texttt{IMFIT} \citep{McMullin2007}. The locations shown in the figure are obtained by averaging the centroid locations derived from individual images in 6.4--14.4 GHz. The lower-frequency sources are excluded because of their lower angular resolution and their sensitivity to lower magnetic field regions, which are presumably located further away from the flare arcade. The uncertainties are determined by the standard deviation of the image source centroids across the frequency range.

Similarly, both the centroid location of the RHESSI 6--12 keV X-ray looptop source and the looptop EUV brightening show a northward trend during the three post-impulsive bursts. In Figure~\ref{fig:source_motion}, the open circles mark the centroids of the RHESSI 6--12 keV source during the main-impulsive phase (red) and the post-impulsive phase bursts (the color code is as same as the microwave circles), with their sizes corresponding to the uncertainties. The centroid location of the X-ray sources and their uncertainties are determined using Detector 3 images made with the \texttt{VIS\_FWDFIT} algorithm. Meanwhile, the star symbols in Figure~\ref{fig:source_motion}(a) represent the locations (pixels) of the brightest SDO/AIA 131 \AA\ emission in the looptop region (within the box labeled ``LT'') at the corresponding times. 
The star symbols are also plotted in the SDO/AIA 131~\AA images in Figure~\ref{fig:source_motion}(d). 

The synchronized northward motion as observed in the looptop region is also observed at the conjugate footpoints of the post-flare arcade. In Figure~\ref{fig:source_motion}(a), the straight ends of the two ribbons are enclosed in the white boxes with ``$R_{n}$'' and ``$R_{s}$''. The crosses mark the brightest points on the flare ribbons in SDO/AIA 1600~\AA images. An enlarged view of the southern and northern ribbons are shown in Figures~\ref{fig:source_motion}(c) and (e), respectively. Similar to the coronal emissions, the conjugate footpoints have a unidirectional northward motion.

During the main-impulsive phase, UV brightening is also observed at the far end of the conjugate ribbons. The locations are shown as red diamonds in Figure~\ref{fig:source_motion}. The southern footpoint brightening is plotted in the inset in Figure~\ref{fig:source_motion}(a) with a larger FOV, in which the full FOV of Figure~\ref{fig:source_motion}(a) is shown as the white box.

The STEREO-A/EUVI images provide further confirmation of the locations of the footpoint brightenings. In Figure~\ref{fig:source_motion}(b), we show the same ribbon brightnenings identified in SDO/AIA 1600 \AA\ images reprojected to STEREO-A/EUVI's viewing perspective (color cross symbols). For the re-projection, we have assumed that the footpoint brightnenings occur at a chromospheric height of 1000 km. Re-projection of the flare arcades seen by SDO/AIA is less straightforward owing to their unknown heights. For illustration purposes, in Figure~\ref{fig:source_motion}(b), we show re-projected flare arcades and the looptop EUV brightenings at the respective times in STEREO-A/EUVI's view, by assuming that all the flare arcades stand vertically above the solar surface.

\begin{figure*}[!ht]
\centering
\includegraphics[width=1.0\textwidth]{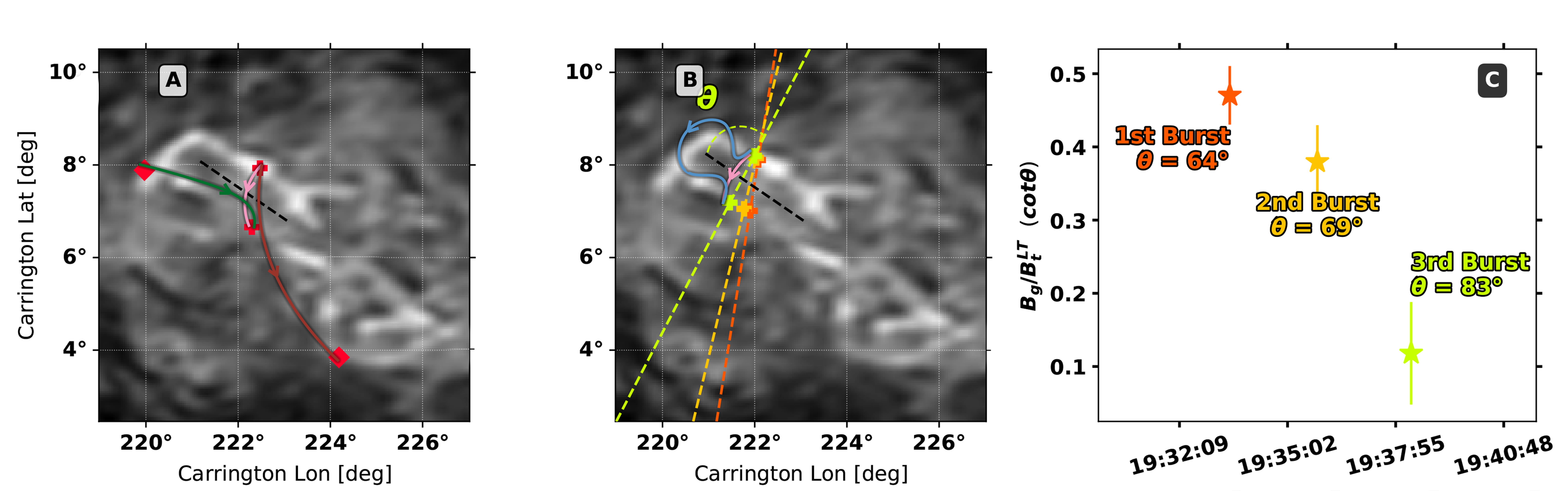}
\caption{\label{fig:shear_angle} (a) Reconnection geometry and the inclination angle during the main impulsive phase. The images are the STEREO-A/EUVI 304~\AA\ image at 19:06 UT re-projected to the heliographic Carrington coordinates. The red crosses show the represented locations of the footpoint brightenings (similar to Figure~\ref{fig:source_motion}(b)). The dashed black line shows the estimated location of the polarity inversion line (PIL). The solid green and red curves demonstrate the pre-reconnection field lines, while the solid pink line shows the post-reconnection arcade. (b) Same as (a), but demonstrating the reconnection geometry (the pre-reconnection field line is demonstrated by the blue curve) and the inclination angle during the three post-impulsive phase bursts. (c) Evolution of the inclination angle $\theta$ during the three post-impulsive bursts and the corresponding normalized guide field estimates $B_g/B_{t}^{LT}$. The color code in (b) and (c) follows Figures~\ref{fig:source_motion}.}
\end{figure*}

The footpoint motion can be used to estimate the inclination angle of the post-reconnection flare arcade with respect to the magnetic polarity inversion line (PIL; \citealt{qiu2010reconnection, qiu2022properties, dahlin2022variability}). However, it is difficult to do so from SDO's viewing perspective owing to the close-to-limb location of the event. To make an estimate, in Figure~\ref{fig:shear_angle}(b), we use the mid-point location between the parallel portion of the two ribbons seen by STEREO-A/EUVI as a proxy for the PIL (shown as a black dashed line). Then, we use straight lines that connect the three pairs of post-impulsive conjugate footpoint brightenings, shown in Figure~\ref{fig:shear_angle}(b) as the orange, yellow, and green dashed lines, respectively, to represent the orientations of the post-reconnection flare arcade. It can be seen that the lines connecting the post-impulsive footpoints become more and more perpendicular with regard to the PIL as the flare progresses, indicating a smaller and smaller shear between the reconnecting magnetic field lines. In order to quantify the evolution of the normalized guide field, following \citealt{qiu2010reconnection, qiu2023role}, we define the inclination angle $\theta$ as the acute angle between the post-reconnection arcade and the PIL. The inclination angle $\theta$ can be used to estimate the normalized guide field: $R_{g} \approx B_{g}/B_{t}^{LT}\approx \cot{\theta}$, where $B_{g}$ and $B_{t}^{LT}$ are the guide field (parallel to the PIL) and the transverse component (perpendicular to the PIL) of the magnetic field at the looptop region, respectively.  
In Figure~\ref{fig:shear_angle}(c), we show the evolution of the inclination angle $\theta$ and $R_g$ during the post-impulsive bursts. The $\theta$ value increases from $64^{\circ}$ during the first burst to $83^{\circ}$ during the last burst, demonstrating a decreasing shear of the flare arcade\footnote{A larger $\theta$ value means a smaller shear. For example, the limiting case of $\theta=90^{\circ}$ corresponds to the case in which the reconnecting field lines are completely anti-parallel to each other, while $\theta=0^{\circ}$ corresponds to purely parallel reconnecting field lines.}. 

In the main-impulsive phase, quantifying the shear and the guide field is less straightforward, as illustrated in Figure~\ref{fig:small_fov}, the geometry is relatively complex. However, we can use the observed ribbon brightenings and coronal loops as constraints. The pre-reconnection loops during the main-impulsive phase, illustrated as the red and green curves in Figure~\ref{fig:shear_angle}(a), correspond to the two loops highlighted in Figure~\ref{fig:small_fov}(b) observed by SDO/AIA 131 \AA. The post-reconnection arcade, displayed as the pink solid arcade in Figure~\ref{fig:shear_angle}(a), corresponds to the arcade in SDO/AIA 131 \AA\ outlined by a pink dashed curve in Figure~\ref{fig:small_fov}(c). 
The outer footpoints of the two pre-reconnection loops located in the hook regions of the ribbons are indicated by a pair of red diamonds in Figure~\ref{fig:shear_angle}(a), and the inner footpoints are shown by a pair of red crosses. In contrast to the post-impulsive phase in which the reconnecting field lines are nearly anti-parallel to each other, the reconnecting loops during the main impulsive phase likely have a very small inclination angle ($\theta\ll 45^{\circ}$), suggestive of a much greater shear or guide field component.   

\subsection{Summary of the Observations}\label{sec:obs_sum}

The main observational phenomena are summarized as follows:

\begin{itemize}
    \item The eruptive flare features three post-impulsive bursts in both microwaves and X-rays. Imaging reveals that the source is located at and above the top of the post-flare arcade.

    \item The power-law index $\delta'$ of the nonthermal electron energy distribution diagnosed using the microwave data at the ALT region shows a notable hardening for later post-impulsive bursts.

    \item The time evolution of the erupting flux rope features multiple episodes of acceleration during the post-impulsive bursts. The last one appears to have the strongest acceleration, perhaps even greater than that during the main-impulsive phase. 

    \item A synchronized northward motion of the microwave/X-ray/EUV looptop source and UV footpoint brightening is observed during the post-impulsive phase. 
    \item The inclination angle of the flare arcades with respect to PIL, inferred by the UV brightening on the flare ribbons, shows an increase for later post-impulsive bursts. In other words, the shear of the post-reconnection flare arcade decreases in time.
    \item The RHESSI observations suggest a softer spectrum during the main-impulsive phase than the post-impulsive phase.
    
\end{itemize}
\section{Discussion and Conclusions}\label{sec:disc}
Our observations suggest a positive correlation between the microwave electron spectral hardness diagnosed by EOVSA microwave imaging spectroscopy and the acceleration of the associated flux rope in multiple microwave/X-ray bursts during the post-impulsive phase of a single eruptive flare event. The observation resembles the widely observed temporal correlation between the CME acceleration and the hard X-ray flux (or SXR derivative) in previous studies \citep[e.g.][]{zhangTemporalRelationshipCoronal2001, Qiu2004, Jing2005, Maricic2007,temmer2008acceleration}. However, most, if not all, of the previous reports were made during the main-impulsive phase of eruptive flares. In contrast, the main-impulsive phase of our event does not correspond to the strongest flux rope acceleration compared to its post-impulsive counterpart. Coincidentally, despite featuring a brighter flare emission in X-rays, the main-impulsive phase has a softer spectrum than the post-impulsive phase. 

\begin{figure*}[!hb]
\centering
\includegraphics[width=0.9\textwidth]{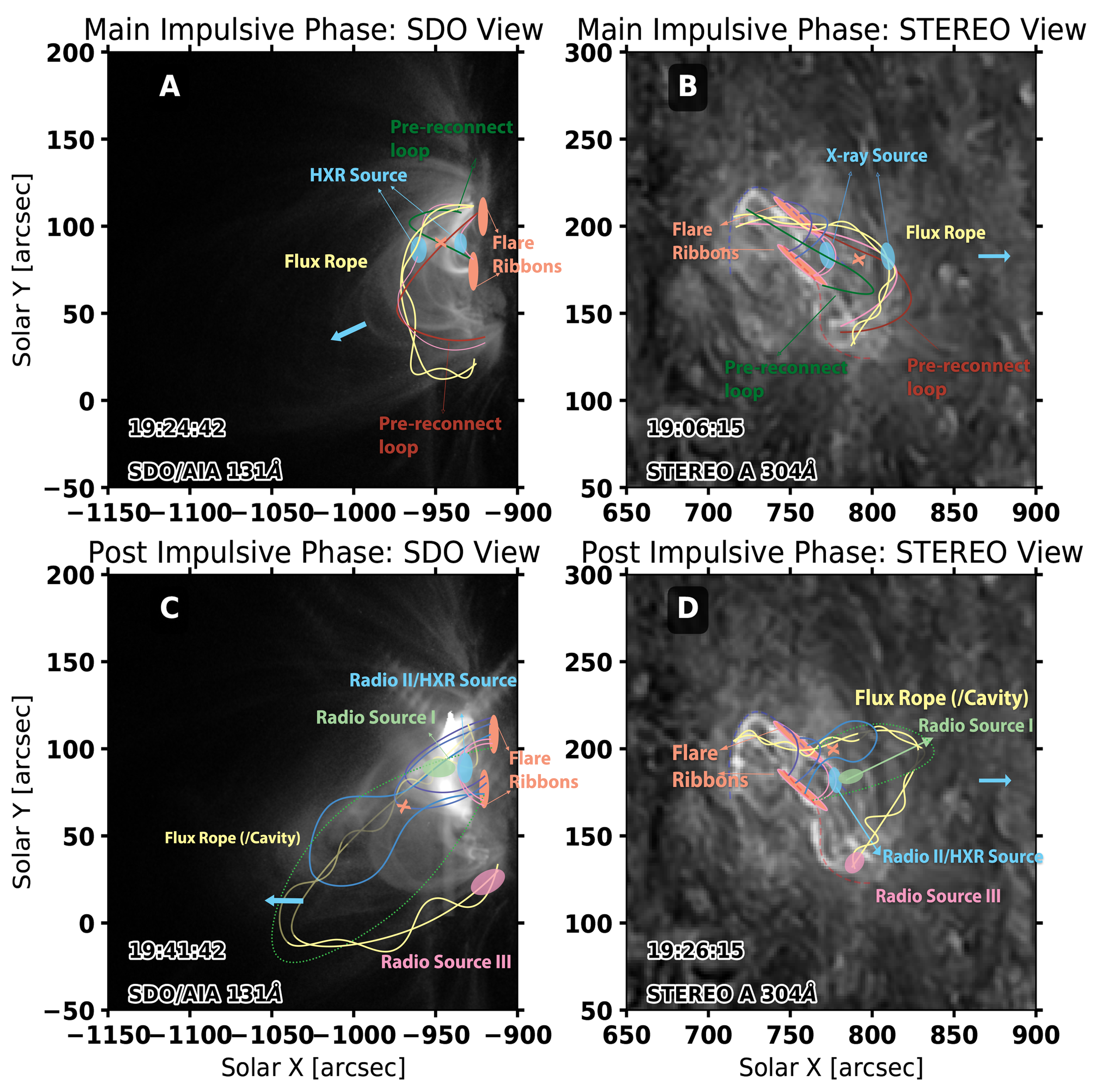}
\caption{\label{fig:cartoon} Schematic cartoon of the flare geometry in the main-impulsive phase (a)--(b) and post-impulsive phase (c)--(d). 
(a) The flare geometry from the viewing perspective of the SDO/AIA during the main-impulsive phase.  The rising flux rope is marked by the twisted yellow curves. Red and green curves represent magnetic field lines reconnecting in a tether-cutting scenario. The reconnection forms a new field line adding to the flux rope (upper pink curve) and a post-flare arcade (lower pink curve). The orange $X$ denotes the reconnection point. (b) Same as (a), but is plotted in the viewing perspective of STEREO-A/EUVI on 304~\AA\ image. 
(c) The flare geometry in the viewing perspective of the SDO/AIA during the post-impulsive phase. The overlying field line around the erupting flux rope cavity is marked by the solid blue curve.  
The EOVSA microwave source (green oval), RHESSI X-ray source (blue oval), post-flare arcades (solid pink curves), and ribbon brightening (orange ovals) are also shown. 
The EOVSA microwave source at the southern footpoint of the erupting flux rope is shown as the pink oval. The background image is the same as in Figure~\ref{fig:big_fov}(a), showing the cross-section of the flux rope cavity. (d) Same as (c), but is plotted in the viewing perspective of STEREO-A/EUVI on 304~\AA\ image. }
\end{figure*}

Figure~\ref{fig:cartoon} places the various observed features into the flare context for the pre- and main-impulsive phase (panels (a) and (b)) and the post-impulsive phase (panels (c) and (d)). During the pre- and main-impulsive phases, two highly sheared magnetic field lines (red and green solid curves in Figure~\ref{fig:cartoon}(a), (b)) reconnect with each other in a ``tether-cutting'' fashion, with their inner footpoints coincide with the ribbon brightenings. After the reconnection, one set of the overlying, highly sheared post-reconnection field lines (upper solid pink line) join the flux rope, and the other set of lower-lying field lines become the bright flare arcade (lower solid pink line). The released energy from the reconnection results in the heating of both the flux rope and the flare arcade, which are observed as the bright coronal structure and the flare arcade seen in EUV (Figure~\ref{fig:small_fov}(d)). 

The heating and associated electron acceleration during this period may be also responsible for the observation of the multiple RHESSI 6--12 keV sources at the bright coronal structure, the looptop, and footpoint (Figure~\ref{fig:mw}(a)). We conclude that the observations during the pre- and main-impulsive phase are broadly consistent with the tether-cutting reconnection scenario (see, e.g., \citealt{chen2014direct}, for similar observations), which may also contribute to the slow rise motion of the magnetic flux rope.

During the post-impulsive phase, the reconnection geometry represents that of the standard scenario for eruptive flares, in which the overlying magnetic fields reconnect below the flux rope. In line with suggestions made in other studies \citep{tripathi2006propagation, Aulanier2012, aulanier2013standard, janvier2013standard, qiu2017elongationa} that expand the essentially two-dimensional standard flare model to three dimensions, our observation of a synchronized northward motion of the microwave/X-ray/EUV looptop source and the UV footpoints favors a scenario in which the flux rope erupts in a zipper-like fashion.  In this scenario, the primary reconnection site moves nearly parallel to the ribbon. The motion can be attributed to asymmetric flux rope eruption, which may be caused by an asymmetric external magnetic confinement  \citep[e.g.][]{tripathi2006propagation,liu2009asymmetric,liu2010motions,zimovets2021quasiperiodica}. Episodic energy release events during the zipper reconnection may be also responsible for the observed multiple episodes of microwave and X-ray bursts. 

We suggest that the observed kinematics of the erupting flux rope is also consistent with the two reconnection scenarios during the pre- to post-impulsive phase of this event. 
The flux rope starts to accelerate after the event enters the main-impulsive phase. Compared to that in the post-impulsive phase (especially bursts \# 2 and 3), the increase in acceleration during the main-impulsive phase is relatively insignificant (Figure~\ref{fig:traj}). In contrast, the acceleration shows a clear increase when entering the post-impulsive phase bursts. In an eruptive flare, the kinematic evolution of the flux rope usually starts with a slow-rise phase followed by an impulsive acceleration phase \citep{zhangTemporalRelationshipCoronal2001}. The slow-rise phase is often found to have an approximately linear height-time profile, and the process is usually attributed to the tether-cutting reconnection scenario \citep{sterling2007new, schrijver2008observations, cheng2020initiation}. By contrast, the fast-rise phase is often attributed to either runaway reconnection \citep{sterling2011insights} or positive feedback from the fast flare reconnection below the flux rope \citep{lin2000effects, cheng2020initiation, liu2021tethercutting, jiang2021}. We suggest that the differences we observe in the main- and post-impulsive phase in the flux rope acceleration are generally similar to those that distinguish the slow- and fast-rise phase in other eruptive flares.

Similar to the previous studies, we attribute the acceleration of the flux rope to the positive feedback from the reconnection occurring in the current sheet trailing the flux rope. However, rather than a dominant and impulsive driver that gives rise to a prominent acceleration period, in our event, the driver may be intermittent during the zipper-like reconnection. The rise of the flux rope is likely asymmetric. It propagates from the active leg on the southwestern side to the anchored leg on the northeastern side (Figure~\ref{fig:cartoon}(c) and (d)), giving rise to the systematic motion of the looptop X-ray, microwave, and EUV sources and the footpoint brightenings. Meanwhile, the intermittent reconnection also drives the multiple acceleration episodes during the post-impulsive phase \citep[see, e.g.,][]{liu2009asymmetric}.

Now we turn our attention to the energization of nonthermal electrons during the main- and post-impulsive phase. As shown in Figure~\ref{fig:rhessi_fitting} and Table~\ref{tab:rhessi_fitting_result}, the power-law index of the observed X-ray spectrum during the main-impulsive phase seems even larger (softer) than that during the post-impulsive phase despite having a much brighter flare emission at $<$ $\sim$20 keV. 
During the impulsive phase, the reconnection between magnetic loops usually has a large shear \citep{Moore1980, Moore2001} (Figure~\ref{fig:shear_angle}(a)).  
Consequently, the guide field is large, which leads to the low productivity of nonthermal electrons during the main impulsive phase \citep{dahlin2016parallel, li2017particle, arnold2021electron, qiu2022properties}. For the post-impulsive phase bursts, we find that the power-law index $\delta'$ of the nonthermal electron energy distribution is harder for later bursts. The hardening of the electron energy spectrum coincides with an increasing acceleration of the flux rope during these post-impulsive bursts. Previous observational and modeling studies have suggested that, when the eruption is well underway, the flux rope acceleration serves as an excellent proxy for the rate of magnetic energy release via reconnection. 
Therefore, we attribute the hardening of the nonthermal electron spectra to an increasing magnetic energy release rate which, in turn, facilitates the acceleration of nonthermal electrons to higher energies. We also note that the inclination angle $\theta$ of the post-flare arcade with regard to the PIL appears to increase throughout the post-impulsive bursts (Figure~\ref{fig:shear_angle}(b)).  Such a change implies a smaller shear of the reconnecting magnetic field and a decreasing guide field. We suggest that such a decreasing guide field component may also contributes to the hardening of the nonthermal electron spectra. 
In summary, we have presented an eruptive flare event that features three post-impulsive X-ray and microwave bursts immediately following its main impulsive phase. We have investigated the relationship between the flux rope acceleration and the electron energization in the context of the flare geometry and its evolution. We have found a positive correlation between the flux rope acceleration and electron energization during the post-impulsive phase bursts, conforming to the standard CME-flare scenario in which positive feedback between flare reconnection and flux rope acceleration is expected. In contrast, such a correlation does not seem to hold during its main impulsive phase. We attribute the lack of flux rope acceleration during the main impulsive phase to the tether-cutting reconnection scenario when the flux rope eruption has not been fully underway. Our observations also suggest a weakening guide field may contribute to the hardening of the nonthermal electron spectrum throughout the main- and post-impulsive phases of the event.

\acknowledgements  

The Expanded Owens Valley Solar Array (EOVSA) was designed, built, and is now operated by the New Jersey Institute of Technology (NJIT) as a community facility. EOVSA operations are supported by NSF grant AGS-2130832 and NASA grant 80NSSC20K0026 to NJIT. This work is primarily supported by NASA grant 80NSSC20K1318 to NJIT. Additionally, it receives support by NSF under grants AGS-1654382, AGS-1954737, AGS-1821294, AST-2108853, AST-2204384, and NASA under grants 80NSSC19K0068, 80NSSC20K0627, 80NSSC20K1282, 80NSSC21K1671, and 80NSSC21K0003 to NJIT. 

\software{SciPy \citep{2020SciPy-NMeth},
          AstroPy \citep{Robitaille2013}, 
          CASA \citep{McMullin2007},
          Lmfit \citep{Newville2016},
          MGN \citep{Morgan2014},
          NumPy \citep{harris2020array},
          SunPy \citep{Community2015}
}

\vspace{1.5in}

\begin{figure*}[!ht]
\begin{interactive}{animation}{figures/movie.mp4}
\plotone{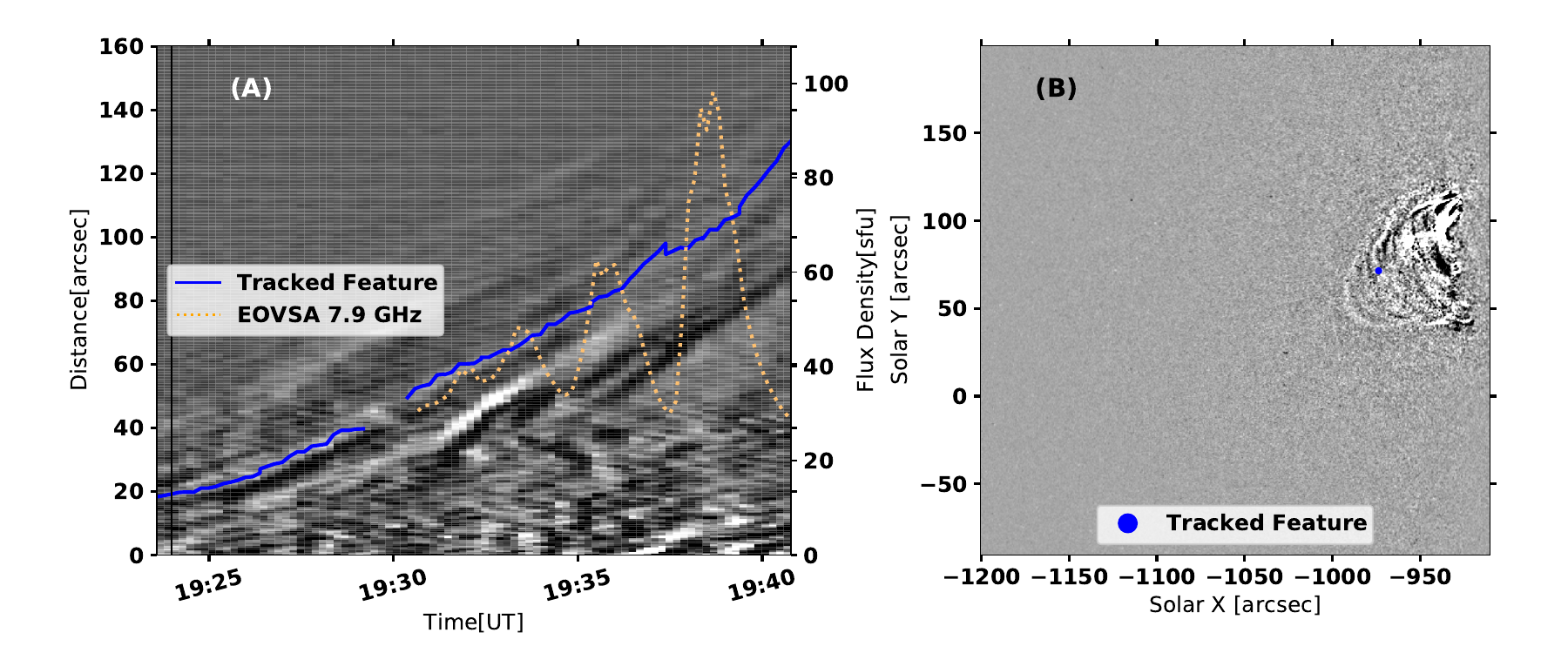}
\end{interactive}
\caption{
This 7-second animation illustrates the ascent of the tracked eruptive feature, identified as the flux rope front, as captured in SDO/AIA 94~\AA\ images. The animation also depicts its trajectory in the corresponding time-distance plot. This animation serves as a supplementary visual aid to Figure~\ref{fig:traj}. (a) Time-distance plot
of SDO/AIA 94~\AA\ background-subtracted images with a cut as shown as the white dashed line in Figure~\ref{fig:big_fov} (g). The orange curve shows the EOVSA microwave flaring-region-integrated light curve at 7.9 GHz. The blue solid line indicates the selected feature for tracking. (b) SDO/AIA 94~\AA\ background-subtracted images. The blue dot is the tracked point along the slit used in (a).}
\label{video:stack_plot}
\end{figure*}

\appendix
Animated Figure~\ref{video:stack_plot} shows the time-evolution of the loop-like feature at the front of the rising magnetic flux rope observed by SDO/AIA 94 \AA\ as well as its trajectory in the time-distance plot.

\bibliography{Wei2021}{}

\begin{thebibliography}{}
\expandafter\ifx\csname natexlab\endcsname\relax\def\natexlab#1{#1}\fi
\providecommand{\url}[1]{\href{#1}{#1}}
\providecommand{\dodoi}[1]{doi:~\href{http://doi.org/#1}{\nolinkurl{#1}}}
\providecommand{\doeprint}[1]{\href{http://ascl.net/#1}{\nolinkurl{http://ascl.net/#1}}}
\providecommand{\doarXiv}[1]{\href{https://arxiv.org/abs/#1}{\nolinkurl{https://arxiv.org/abs/#1}}}

\bibitem[{Antiochos {et~al.}(1999)Antiochos, DeVore, \&
  Klimchuk}]{antiochos1999model}
Antiochos, S.~K., DeVore, C.~R., \& Klimchuk, J.~A. 1999, The Astrophysical
  Journal, 510, 485, \dodoi{10.1086/306563}

\bibitem[{Arnold {et~al.}(2021)Arnold, Drake, Swisdak, Guo, Dahlin, Chen,
  Fleishman, Glesener, Kontar, Phan, \& Shen}]{arnold2021electron}
Arnold, H., Drake, J.~F., Swisdak, M., {et~al.} 2021, Physical Review Letters,
  126, 135101, \dodoi{10.1103/PhysRevLett.126.135101}

\bibitem[{Aulanier {et~al.}(2013)Aulanier, D{\'e}moulin, Schrijver, Janvier,
  Pariat, \& Schmieder}]{aulanier2013standard}
Aulanier, G., D{\'e}moulin, P., Schrijver, C.~J., {et~al.} 2013, Astronomy \&
  Astrophysics, 549, A66, \dodoi{10.1051/0004-6361/201220406}

\bibitem[{Aulanier {et~al.}(2012)Aulanier, Janvier, \&
  Schmieder}]{Aulanier2012}
Aulanier, G., Janvier, M., \& Schmieder, B. 2012, Astronomy {\&} Astrophysics,
  543, A110, \dodoi{10.1051/0004-6361/201219311}

\bibitem[{Brueckner {et~al.}(1995)Brueckner, Howard, Koomen, Korendyke,
  Michels, Moses, Socker, Dere, Lamy, Llebaria, Bout, Schwenn, Simnett,
  Bedford, \& Eyles}]{brueckner1995large}
Brueckner, G.~E., Howard, R.~A., Koomen, M.~J., {et~al.} 1995, in The {{SOHO
  Mission}}, ed. B.~Fleck, V.~Domingo, \& A.~Poland ({Dordrecht}: {Springer
  Netherlands}), 357--402, \dodoi{10.1007/978-94-009-0191-9_10}

\bibitem[{Chen {et~al.}(2020{\natexlab{a}})Chen, Yu, Reeves, \&
  Gary}]{chen2020microwave}
Chen, B., Yu, S., Reeves, K.~K., \& Gary, D.~E. 2020{\natexlab{a}}, The
  Astrophysical Journal Letters, 895, L50, \dodoi{10.3847/2041-8213/ab901a}

\bibitem[{Chen {et~al.}(2020{\natexlab{b}})Chen, Shen, Gary, Reeves, Fleishman,
  Yu, Guo, Krucker, Lin, Nita, \& Kong}]{chen2020measurement}
Chen, B., Shen, C., Gary, D.~E., {et~al.} 2020{\natexlab{b}}, Nature Astronomy,
  4, 1140, \dodoi{10.1038/s41550-020-1147-7}

\bibitem[{Chen {et~al.}(2014)Chen, Zhang, Cheng, Ma, Yang, \&
  Li}]{chen2014direct}
Chen, H., Zhang, J., Cheng, X., {et~al.} 2014, The Astrophysical Journal
  Letters, 797, L15, \dodoi{10.1088/2041-8205/797/2/L15}

\bibitem[{Chen(2011)}]{chen2011coronal}
Chen, P.~F. 2011, Living Reviews in Solar Physics, 8, 1,
  \dodoi{10.12942/lrsp-2011-1}

\bibitem[{Cheng {et~al.}(2003)Cheng, Ren, Choe, \& Moon}]{cheng2003fluxa}
Cheng, C.~Z., Ren, Y., Choe, G.~S., \& Moon, Y.-J. 2003, The Astrophysical
  Journal, 596, 1341, \dodoi{10.1086/378170}

\bibitem[{Cheng {et~al.}(2020)Cheng, Zhang, Kliem, T{\"o}r{\"o}k, Xing, Zhou,
  Inhester, \& Ding}]{cheng2020initiation}
Cheng, X., Zhang, J., Kliem, B., {et~al.} 2020, The Astrophysical Journal, 894,
  85, \dodoi{10.3847/1538-4357/ab886a}

\bibitem[{Community {et~al.}(2015)Community, Mumford, Christe,
  P{\'{e}}rez-Su{\'{a}}rez, Ireland, Shih, Inglis, Liedtke, Hewett, Mayer,
  Hughitt, Freij, Meszaros, Bennett, Malocha, Evans, Agrawal, Leonard,
  Robitaille, Mampaey, Campos-Rozo, \& Kirk}]{Community2015}
Community, T.~S., Mumford, S.~J., Christe, S., {et~al.} 2015, Computational
  Science {\&} Discovery, 8, 014009, \dodoi{10.1088/1749-4699/8/1/014009}

\bibitem[{Dahlin {et~al.}(2022)Dahlin, Antiochos, Qiu, \&
  DeVore}]{dahlin2022variability}
Dahlin, J.~T., Antiochos, S.~K., Qiu, J., \& DeVore, C.~R. 2022, The
  Astrophysical Journal, 932, 94, \dodoi{10.3847/1538-4357/ac6e3d}

\bibitem[{Dahlin {et~al.}(2014)Dahlin, Drake, \&
  Swisdak}]{dahlin2014mechanisms}
Dahlin, J.~T., Drake, J.~F., \& Swisdak, M. 2014, Physics of Plasmas, 21,
  092304, \dodoi{10.1063/1.4894484}

\bibitem[{Dahlin {et~al.}(2016)Dahlin, Drake, \& Swisdak}]{dahlin2016parallel}
---. 2016, Physics of Plasmas, 23, 120704, \dodoi{10.1063/1.4972082}

\bibitem[{De~Pontieu {et~al.}(2014)De~Pontieu, Title, Lemen, Kushner, Akin,
  Allard, Berger, Boerner, Cheung, Chou, Drake, Duncan, Freeland, Heyman,
  Hoffman, Hurlburt, Lindgren, Mathur, Rehse, Sabolish, Seguin, Schrijver,
  Tarbell, W{\"u}lser, Wolfson, Yanari, Mudge, {Nguyen-Phuc}, Timmons, {van
  Bezooijen}, Weingrod, Brookner, Butcher, Dougherty, Eder, Knagenhjelm,
  Larsen, Mansir, Phan, Boyle, Cheimets, DeLuca, Golub, Gates, Hertz, McKillop,
  Park, Perry, Podgorski, Reeves, Saar, Testa, Tian, Weber, Dunn, Eccles,
  Jaeggli, Kankelborg, Mashburn, Pust, Springer, Carvalho, Kleint, Marmie,
  Mazmanian, Pereira, Sawyer, Strong, Worden, Carlsson, Hansteen, Leenaarts,
  Wiesmann, Aloise, Chu, Bush, Scherrer, Brekke, {Martinez-Sykora}, Lites,
  McIntosh, Uitenbroek, Okamoto, Gummin, Auker, Jerram, Pool, \&
  Waltham}]{depontieu2014interface}
De~Pontieu, B., Title, A.~M., Lemen, J.~R., {et~al.} 2014, Solar Physics, 289,
  2733, \dodoi{10.1007/s11207-014-0485-y}

\bibitem[{Dulk(1985)}]{dulk1985radio}
Dulk, G.~A. 1985, Annual Review of Astronomy and Astrophysics, 23, 169,
  \dodoi{10.1146/annurev.aa.23.090185.001125}

\bibitem[{Elmore {et~al.}(2003)Elmore, Burkepile, Darnell, Lecinski, \&
  Stanger}]{elmore2003calibration}
Elmore, D.~F., Burkepile, J.~T., Darnell, J.~A., Lecinski, A.~R., \& Stanger,
  A.~L. 2003, in Polarimetry in {{Astronomy}}, Vol. 4843 ({SPIE}), 66--75,
  \dodoi{10.1117/12.459279}

\bibitem[{Fleishman \& Kuznetsov(2010)}]{Fleishman2010}
Fleishman, G.~D., \& Kuznetsov, A.~A. 2010, The Astrophysical Journal, 721,
  1127, \dodoi{10.1088/0004-637X/721/2/1127}

\bibitem[{Freeland \& Handy(1998)}]{freeland1998data}
Freeland, S., \& Handy, B. 1998, Solar Physics, 182, 497,
  \dodoi{10.1023/A:1005038224881}

\bibitem[{Gary {et~al.}(2018)Gary, Chen, Dennis, Fleishman, Hurford, Krucker,
  McTiernan, Nita, Shih, White, \& Yu}]{Gary2018}
Gary, D.~E., Chen, B., Dennis, B.~R., {et~al.} 2018, The Astrophysical Journal,
  863, 83, \dodoi{10.3847/1538-4357/aad0ef}

\bibitem[{Gieseler {et~al.}(2023)Gieseler, Dresing, Palmroos, {Freiherr von
  Forstner}, Price, Vainio, Kouloumvakos, {Rodr{\'i}guez-Garc{\'i}a}, Trotta,
  G{\'e}not, Masson, Roth, \& Veronig}]{gieseler2023solarmach}
Gieseler, J., Dresing, N., Palmroos, C., {et~al.} 2023, Frontiers in Astronomy
  and Space Sciences, 9

\bibitem[{Gryciuk {et~al.}(2017)Gryciuk, Siarkowski, Sylwester, Gburek,
  Podgorski, Kepa, Sylwester, \& Mrozek}]{gryciuk2017flare}
Gryciuk, M., Siarkowski, M., Sylwester, J., {et~al.} 2017, Solar Physics, 292,
  77, \dodoi{10.1007/s11207-017-1101-8}

\bibitem[{Hannah \& Kontar(2012)}]{hannah2012differential}
Hannah, I.~G., \& Kontar, E.~P. 2012, Astronomy \& Astrophysics, 539, A146,
  \dodoi{10.1051/0004-6361/201117576}

\bibitem[{Harris {et~al.}(2020)Harris, Millman, van~der Walt, Gommers,
  Virtanen, Cournapeau, Wieser, Taylor, Berg, Smith, Kern, Picus, Hoyer, van
  Kerkwijk, Brett, Haldane, del R{\'{i}}o, Wiebe, Peterson,
  G{\'{e}}rard-Marchant, Sheppard, Reddy, Weckesser, Abbasi, Gohlke, \&
  Oliphant}]{harris2020array}
Harris, C.~R., Millman, K.~J., van~der Walt, S.~J., {et~al.} 2020, Nature, 585,
  357, \dodoi{10.1038/s41586-020-2649-2}

\bibitem[{{Haw} {et~al.}(2018){Haw}, {Wongwaitayakornkul}, {Li}, \&
  {Bellan}}]{2018ApJ...862L..15H}
{Haw}, M.~A., {Wongwaitayakornkul}, P., {Li}, H., \& {Bellan}, P.~M. 2018,
  \apjl, 862, L15, \dodoi{10.3847/2041-8213/aad33c}

\bibitem[{Howard {et~al.}(2017)Howard, DeForest, Schneck, \&
  Alden}]{howard2017challenginga}
Howard, T.~A., DeForest, C.~E., Schneck, U.~G., \& Alden, C.~R. 2017, The
  Astrophysical Journal, 834, 86, \dodoi{10.3847/1538-4357/834/1/86}

\bibitem[{Hu {et~al.}(2014)Hu, Qiu, Dasgupta, Khare, \&
  Webb}]{hu2014structures}
Hu, Q., Qiu, J., Dasgupta, B., Khare, A., \& Webb, G.~M. 2014, The
  Astrophysical Journal, 793, 53, \dodoi{10.1088/0004-637X/793/1/53}

\bibitem[{Hurford {et~al.}(2002)Hurford, Schmahl, Schwartz, Conway, Aschwanden,
  Csillaghy, Dennis, Johns-Krull, Krucker, Lin, McTiernan, Metcalf, Sato, \&
  Smith}]{Hurford2002}
Hurford, G.~J., Schmahl, E.~J., Schwartz, R.~A., {et~al.} 2002, Solar Physics,
  210, 61, \dodoi{10.1023/A:1022436213688}

\bibitem[{Illing \& Hundhausen(1985)}]{illing1985observation}
Illing, R. M.~E., \& Hundhausen, A.~J. 1985, Journal of Geophysical Research:
  Space Physics, 90, 275, \dodoi{10.1029/JA090iA01p00275}

\bibitem[{Janvier {et~al.}(2013)Janvier, Aulanier, Pariat, \&
  D{\'e}moulin}]{janvier2013standard}
Janvier, M., Aulanier, G., Pariat, E., \& D{\'e}moulin, P. 2013, Astronomy \&
  Astrophysics, 555, A77, \dodoi{10.1051/0004-6361/201321164}

\bibitem[{Jiang {et~al.}(2021)Jiang, Feng, Liu, Yan, Hu, Moore, Duan, Cui, Zuo,
  Wang, \& Wei}]{jiang2021}
Jiang, C., Feng, X., Liu, R., {et~al.} 2021, Nature Astronomy, 1,
  \dodoi{10.1038/s41550-021-01414-z}

\bibitem[{Jing {et~al.}(2005)Jing, Qiu, Lin, Qu, Xu, \& Wang}]{Jing2005}
Jing, J., Qiu, J., Lin, J., {et~al.} 2005, The Astrophysical Journal, 620,
  1085, \dodoi{10.1086/427165}

\bibitem[{Kaiser {et~al.}(2008)Kaiser, Kucera, Davila, St.~Cyr, Guhathakurta,
  \& Christian}]{kaiser2008stereo}
Kaiser, M.~L., Kucera, T.~A., Davila, J.~M., {et~al.} 2008, Space Science
  Reviews, 136, 5, \dodoi{10.1007/s11214-007-9277-0}

\bibitem[{Karpen {et~al.}(2012)Karpen, Antiochos, \&
  DeVore}]{karpen2012mechanisms}
Karpen, J.~T., Antiochos, S.~K., \& DeVore, C.~R. 2012, The Astrophysical
  Journal, 760, 81, \dodoi{10.1088/0004-637X/760/1/81}

\bibitem[{Kou {et~al.}(2022)Kou, Cheng, Wang, Yu, Chen, Kontar, \&
  Ding}]{kou2022microwave}
Kou, Y., Cheng, X., Wang, Y., {et~al.} 2022, Nature Communications, 13, 7680,
  \dodoi{10.1038/s41467-022-35377-0}

\bibitem[{Lemen {et~al.}(2012)Lemen, Title, Akin, Boerner, Chou, Drake, Duncan,
  Edwards, Friedlaender, Heyman, Hurlburt, Katz, Kushner, Levay, Lindgren,
  Mathur, McFeaters, Mitchell, Rehse, Schrijver, Springer, Stern, Tarbell,
  Wuelser, Wolfson, Yanari, Bookbinder, Cheimets, Caldwell, Deluca, Gates,
  Golub, Park, Podgorski, Bush, Scherrer, Gummin, Smith, Auker, Jerram, Pool,
  Soufli, Windt, Beardsley, Clapp, Lang, \& Waltham}]{lemen2012}
Lemen, J.~R., Title, A.~M., Akin, D.~J., {et~al.} 2012, Solar Physics, 275, 17,
  \dodoi{10.1007/s11207-011-9776-8}

\bibitem[{Li {et~al.}(2017)Li, Guo, Li, \& Li}]{li2017particle}
Li, X., Guo, F., Li, H., \& Li, G. 2017, The Astrophysical Journal, 843, 21,
  \dodoi{10.3847/1538-4357/aa745e}

\bibitem[{Lin \& Forbes(2000)}]{lin2000effects}
Lin, J., \& Forbes, T.~G. 2000, Journal of Geophysical Research: Space Physics,
  105, 2375, \dodoi{10.1029/1999JA900477}

\bibitem[{Lin {et~al.}(2005)Lin, Ko, Sui, Raymond, Stenborg, Jiang, Zhao, \&
  Mancuso}]{lin2005direct}
Lin, J., Ko, Y.-K., Sui, L., {et~al.} 2005, The Astrophysical Journal, 622,
  1251, \dodoi{10.1086/428110}

\bibitem[{Lin {et~al.}(2002)Lin, Dennis, Hurford, Smith, Zehnder, Harvey,
  Curtis, Pankow, Turin, Bester, Csillaghy, Lewis, Madden, {van Beek}, Appleby,
  Raudorf, McTiernan, Ramaty, Schmahl, Schwartz, Krucker, Abiad, Quinn, Berg,
  Hashii, Sterling, Jackson, Pratt, Campbell, Malone, Landis,
  {Barrington-Leigh}, {Slassi-Sennou}, Cork, Clark, Amato, Orwig, Boyle, Banks,
  Shirey, Tolbert, Zarro, Snow, Thomsen, Henneck, Mchedlishvili, Ming, Fivian,
  Jordan, Wanner, Crubb, Preble, Matranga, Benz, Hudson, Canfield, Holman,
  Crannell, Kosugi, Emslie, Vilmer, Brown, {Johns-Krull}, Aschwanden, Metcalf,
  \& Conway}]{lin2002}
Lin, R., Dennis, B., Hurford, G., {et~al.} 2002, Solar Physics, 210, 3,
  \dodoi{10.1023/A:1022428818870}

\bibitem[{Liu {et~al.}(2010)Liu, Lee, Jing, Liu, Deng, \&
  Wang}]{liu2010motions}
Liu, C., Lee, J., Jing, J., {et~al.} 2010, The Astrophysical Journal Letters,
  721, L193, \dodoi{10.1088/2041-8205/721/2/L193}

\bibitem[{Liu \& Wang(2009)}]{liu2009reconnection}
Liu, C., \& Wang, H. 2009, The Astrophysical Journal, 696, L27,
  \dodoi{10.1088/0004-637X/696/1/L27}

\bibitem[{Liu {et~al.}(2009)Liu, Alexander, \& Gilbert}]{liu2009asymmetric}
Liu, R., Alexander, D., \& Gilbert, H.~R. 2009, The Astrophysical Journal, 691,
  1079, \dodoi{10.1088/0004-637X/691/2/1079}

\bibitem[{Liu \& Su(2021)}]{liu2021tethercutting}
Liu, T., \& Su, Y. 2021, The Astrophysical Journal, 915, 55,
  \dodoi{10.3847/1538-4357/ac013a}

\bibitem[{Mari{\v c}i{\'c} {et~al.}(2007)Mari{\v c}i{\'c}, Vr{\v s}nak,
  Stanger, Veronig, Temmer, \& Ro{\v s}a}]{Maricic2007}
Mari{\v c}i{\'c}, D., Vr{\v s}nak, B., Stanger, A.~L., {et~al.} 2007, Solar
  Physics, 241, 99, \dodoi{10.1007/s11207-007-0291-x}

\bibitem[{McMullin {et~al.}(2007)McMullin, Waters, Schiebel, Young, \&
  Golap}]{McMullin2007}
McMullin, J.~P., Waters, B., Schiebel, D., Young, W., \& Golap, K. 2007,
  Astronomical Data Analysis Software and Systems XVI, 376, 127.
\newblock \url{https://ui.adsabs.harvard.edu/abs/2007ASPC..376..127M/abstract}

\bibitem[{Moore \& Labonte(1980)}]{Moore1980}
Moore, R.~L., \& Labonte, B.~J. 1980, IAUS, 91, 207.
\newblock \url{https://ui.adsabs.harvard.edu/abs/1980IAUS...91..207M/abstract}

\bibitem[{Moore {et~al.}(2001)Moore, Sterling, Hudson, \& Lemen}]{Moore2001}
Moore, R.~L., Sterling, A.~C., Hudson, H.~S., \& Lemen, J.~R. 2001, The
  Astrophysical Journal, 552, 833, \dodoi{10.1086/320559}

\bibitem[{Morgan \& Druckm{\"{u}}ller(2014)}]{Morgan2014}
Morgan, H., \& Druckm{\"{u}}ller, M. 2014, Solar Physics, 289, 2945,
  \dodoi{10.1007/s11207-014-0523-9}

\bibitem[{Naus {et~al.}(2022)Naus, Qiu, DeVore, Antiochos, Dahlin, Drake, \&
  Swisdak}]{naus2022correlated}
Naus, S.~J., Qiu, J., DeVore, C.~R., {et~al.} 2022, The Astrophysical Journal,
  926, 218, \dodoi{10.3847/1538-4357/ac4028}

\bibitem[{Neupert(1968)}]{neupert1968comparison}
Neupert, W.~M. 1968, The Astrophysical Journal, 153, L59,
  \dodoi{10.1086/180220}

\bibitem[{Newville {et~al.}(2016)Newville, Stensitzki, Allen, Rawlik,
  Ingargiola, Nelson, Newville, Stensitzki, Allen, Rawlik, Ingargiola, \&
  Nelson}]{Newville2016}
Newville, M., Stensitzki, T., Allen, D.~B., {et~al.} 2016, ascl, ascl:1606.014.
\newblock \url{https://ui.adsabs.harvard.edu/abs/2016ascl.soft06014N/abstract}

\bibitem[{Parker(1957)}]{parker1957sweet}
Parker, E.~N. 1957, Journal of Geophysical Research (1896-1977), 62, 509,
  \dodoi{10.1029/JZ062i004p00509}

\bibitem[{Pesnell {et~al.}(2012)Pesnell, Thompson, \& Chamberlin}]{pesnell2012}
Pesnell, W., Thompson, B., \& Chamberlin, P. 2012, Solar Physics, 275, 3,
  \dodoi{10.1007/s11207-011-9841-3}

\bibitem[{Pritchett \& Coroniti(2004)}]{pritchett2004threedimensional}
Pritchett, P.~L., \& Coroniti, F.~V. 2004, Journal of Geophysical Research:
  Space Physics, 109, \dodoi{10.1029/2003JA009999}

\bibitem[{Qiu \& Cheng(2022)}]{qiu2022properties}
Qiu, J., \& Cheng, J. 2022, Solar Physics, 297, 80,
  \dodoi{10.1007/s11207-022-02003-7}

\bibitem[{Qiu {et~al.}(2010)Qiu, Liu, Hill, \&
  Kazachenko}]{qiu2010reconnection}
Qiu, J., Liu, W., Hill, N., \& Kazachenko, M. 2010, The Astrophysical Journal,
  725, 319, \dodoi{10.1088/0004-637X/725/1/319}

\bibitem[{Qiu {et~al.}(2017)Qiu, Longcope, Cassak, \&
  Priest}]{qiu2017elongationa}
Qiu, J., Longcope, D.~W., Cassak, P.~A., \& Priest, E.~R. 2017, The
  Astrophysical Journal, 838, 17, \dodoi{10.3847/1538-4357/aa6341}

\bibitem[{Qiu {et~al.}(2004)Qiu, Wang, Cheng, \& Gary}]{Qiu2004}
Qiu, J., Wang, H., Cheng, C.~Z., \& Gary, D.~E. 2004, The Astrophysical
  Journal, 604, 900, \dodoi{10.1086/382122}

\bibitem[{Qiu {et~al.}(2023)Qiu, Alaoui, Antiochos, Dahlin, Swisdak, Drake,
  Robison, DeVore, \& Uritsky}]{qiu2023role}
Qiu, J., Alaoui, M., Antiochos, S.~K., {et~al.} 2023, The Astrophysical
  Journal, 955, 34, \dodoi{10.3847/1538-4357/acebeb}

\bibitem[{Reeves(2006)}]{Reeves2006}
Reeves, K.~K. 2006, The Astrophysical Journal, 644, 592, \dodoi{10.1086/503352}

\bibitem[{Reeves \& Moats(2010)}]{reeves2010relating}
Reeves, K.~K., \& Moats, S.~J. 2010, The Astrophysical Journal, 712, 429,
  \dodoi{10.1088/0004-637X/712/1/429}

\bibitem[{Robitaille {et~al.}(2013)Robitaille, Tollerud, Greenfield,
  Droettboom, Bray, Aldcroft, Davis, Ginsburg, Price-Whelan, Kerzendorf,
  Conley, Crighton, Barbary, Muna, Ferguson, Grollier, Parikh, Nair,
  G{\"{u}}nther, Deil, Woillez, Conseil, Kramer, Turner, Singer, Fox, Weaver,
  Zabalza, Edwards, {Azalee Bostroem}, Burke, Casey, Crawford, Dencheva, Ely,
  Jenness, Labrie, Lim, Pierfederici, Pontzen, Ptak, Refsdal, Servillat, \&
  Streicher}]{Robitaille2013}
Robitaille, T.~P., Tollerud, E.~J., Greenfield, P., {et~al.} 2013, Astronomy
  {\&} Astrophysics, 558, A33, \dodoi{10.1051/0004-6361/201322068}

\bibitem[{Schrijver {et~al.}(2008)Schrijver, Elmore, Kliem, T{\"o}r{\"o}k, \&
  Title}]{schrijver2008observations}
Schrijver, C.~J., Elmore, C., Kliem, B., T{\"o}r{\"o}k, T., \& Title, A.~M.
  2008, The Astrophysical Journal, 674, 586, \dodoi{10.1086/524294}

\bibitem[{Song {et~al.}(2019)Song, Zhang, Cheng, Li, Tang, Wang, Zheng, \&
  Chen}]{song2019nature}
Song, H.~Q., Zhang, J., Cheng, X., {et~al.} 2019, The Astrophysical Journal,
  883, 43, \dodoi{10.3847/1538-4357/ab304c}

\bibitem[{Sterling {et~al.}(2007)Sterling, Harra, \& Moore}]{sterling2007new}
Sterling, A.~C., Harra, L.~K., \& Moore, R.~L. 2007, The Astrophysical Journal,
  669, 1359, \dodoi{10.1086/520829}

\bibitem[{Sterling {et~al.}(2011)Sterling, Moore, \&
  Freeland}]{sterling2011insights}
Sterling, A.~C., Moore, R.~L., \& Freeland, S.~L. 2011, The Astrophysical
  Journal Letters, 731, L3, \dodoi{10.1088/2041-8205/731/1/L3}

\bibitem[{Temmer {et~al.}(2007)Temmer, Veronig, Vr{\v s}nak, \&
  Miklenic}]{temmer2007energy}
Temmer, M., Veronig, A.~M., Vr{\v s}nak, B., \& Miklenic, C. 2007, The
  Astrophysical Journal, 654, 665, \dodoi{10.1086/509634}

\bibitem[{Temmer {et~al.}(2008)Temmer, Veronig, Vr{\v s}nak, Ryb{\'a}k,
  G{\"o}m{\"o}ry, Stoiser, \& Mari{\v c}i{\'c}}]{temmer2008acceleration}
Temmer, M., Veronig, A.~M., Vr{\v s}nak, B., {et~al.} 2008, The Astrophysical
  Journal, 673, L95, \dodoi{10.1086/527414}

\bibitem[{Tripathi {et~al.}(2006)Tripathi, Isobe, \&
  Mason}]{tripathi2006propagation}
Tripathi, D., Isobe, H., \& Mason, H.~E. 2006, Astronomy \& Astrophysics, 453,
  1111, \dodoi{10.1051/0004-6361:20064993}

\bibitem[{Veronig {et~al.}(2018)Veronig, Podladchikova, Dissauer, Temmer,
  Seaton, Long, Guo, Vr{\v s}nak, Harra, \& Kliem}]{veronig2018genesis}
Veronig, A.~M., Podladchikova, T., Dissauer, K., {et~al.} 2018, The
  Astrophysical Journal, 868, 107, \dodoi{10.3847/1538-4357/aaeac5}

\bibitem[{Virtanen {et~al.}(2020)Virtanen, Gommers, Oliphant, Haberland, Reddy,
  Cournapeau, Burovski, Peterson, Weckesser, Bright, {van der Walt}, Brett,
  Wilson, Millman, Mayorov, Nelson, Jones, Kern, Larson, Carey, Polat, Feng,
  Moore, {VanderPlas}, Laxalde, Perktold, Cimrman, Henriksen, Quintero, Harris,
  Archibald, Ribeiro, Pedregosa, {van Mulbregt}, \& {SciPy 1.0
  Contributors}}]{2020SciPy-NMeth}
Virtanen, P., Gommers, R., Oliphant, T.~E., {et~al.} 2020, Nature Methods, 17,
  261, \dodoi{10.1038/s41592-019-0686-2}

\bibitem[{Vourlidas {et~al.}(2013)Vourlidas, Lynch, Howard, \&
  Li}]{vourlidas2013how}
Vourlidas, A., Lynch, B.~J., Howard, R.~A., \& Li, Y. 2013, Solar Physics, 284,
  179, \dodoi{10.1007/s11207-012-0084-8}

\bibitem[{Wang {et~al.}(2007)Wang, Sui, \& Qiu}]{wang2007direct}
Wang, T., Sui, L., \& Qiu, J. 2007, The Astrophysical Journal, 661, L207,
  \dodoi{10.1086/519004}

\bibitem[{Welsch(2018)}]{welsch2018flux}
Welsch, B.~T. 2018, Solar Physics, 293, 113, \dodoi{10.1007/s11207-018-1329-y}

\bibitem[{Wuelser {et~al.}(2004)Wuelser, Lemen, Tarbell, Wolfson, Cannon,
  Carpenter, Duncan, Gradwohl, Meyer, Moore, Navarro, Pearson, Rossi, Springer,
  Howard, Moses, Newmark, Delaboudiniere, Artzner, Auchere, Bougnet, Bouyries,
  Bridou, Clotaire, Colas, Delmotte, Jerome, Lamare, Mercier, Mullot, Ravet,
  Song, Bothmer, \& Deutsch}]{wuelser2004euvi}
Wuelser, J.-P., Lemen, J.~R., Tarbell, T.~D., {et~al.} 2004, in Telescopes and
  {{Instrumentation}} for {{Solar Astrophysics}}, Vol. 5171 ({SPIE}), 111--122,
  \dodoi{10.1117/12.506877}

\bibitem[{Xue {et~al.}(2016)Xue, Yan, Cheng, Yang, Su, Kliem, Zhang, Liu, Bi,
  Xiang, Yang, \& Zhao}]{xue2016observing}
Xue, Z., Yan, X., Cheng, X., {et~al.} 2016, Nature Communications, 7, 11837,
  \dodoi{10.1038/ncomms11837}

\bibitem[{Yu {et~al.}(2020)Yu, Chen, Reeves, Gary, Musset, Fleishman, Nita, \&
  Glesener}]{Yu2020}
Yu, S., Chen, B., Reeves, K.~K., {et~al.} 2020, The Astrophysical Journal, 900,
  17, \dodoi{10.3847/1538-4357/aba8a6}

\bibitem[{Zhang {et~al.}(2001)Zhang, Dere, Howard, Kundu, \&
  White}]{zhangTemporalRelationshipCoronal2001}
Zhang, J., Dere, K.~P., Howard, R.~A., Kundu, M.~R., \& White, S.~M. 2001, The
  Astrophysical Journal, 559, 452, \dodoi{10.1086/322405}

\bibitem[{Zhu {et~al.}(2020)Zhu, Qiu, Liewer, Vourlidas, Spiegel, \&
  Hu}]{zhu2020how}
Zhu, C., Qiu, J., Liewer, P., {et~al.} 2020, The Astrophysical Journal, 893,
  141, \dodoi{10.3847/1538-4357/ab838a}

\bibitem[{Zimovets {et~al.}(2021)Zimovets, Sharykin, \&
  Myshyakov}]{zimovets2021quasiperiodica}
Zimovets, I., Sharykin, I., \& Myshyakov, I. 2021, Solar Physics, 296, 188,
  \dodoi{10.1007/s11207-021-01936-9}

\end{thebibliography}
\bibliographystyle{aasjournal}


\end{document}